\documentclass[smallextended]{svjour3} 
\usepackage[english]{babel}
\usepackage[utf8]{inputenc}
\usepackage{pgf,tikz,pgfplots}
\pgfplotsset{compat=1.15}
\usetikzlibrary{arrows}

\usepackage{enumerate}
\usepackage{pifont}
\usepackage{grffile}
\usepackage[normalem]{ulem}
\usepackage{amsfonts}
 \usepackage{setspace}
 \expandafter\let\csname equation*\endcsname\relax
\expandafter\let\csname endequation*\endcsname\relax
\usepackage[utf8]{inputenc}
\usepackage{BOONDOX-cal}
\usepackage{bbm}
\usepackage{epsfig}
\usepackage{color}
\usepackage{bm}
\usepackage{float}
\usepackage{physics}
\usepackage{bigints}
\usepackage{amssymb}
\usepackage{subfigure}
\usepackage{setspace}
\usepackage{textcomp}
\usepackage[utf8]{inputenc}
\usepackage{textcomp}
\usepackage{amsmath}
\usepackage{esint}
\usepackage{xcolor}
\usepackage{subfigure}
\usepackage{float}
\usepackage{soul}
\usepackage{stackrel}
\usepackage{color}
\usepackage[bottom]{footmisc}
\usepackage[
colorlinks=true, 
pdfstartview=FitV, 
linkcolor=blue, 
citecolor=magenta, 
urlcolor=blue, 
bookmarks=true,
bookmarksnumbered=true,
pdftitle={},
pdfauthor={}
]{hyperref}
\usepackage{epsfig,amsopn}
\usepackage{graphicx}

\definecolor{pink}{rgb}{1,0,0.6}
\definecolor{dgreen}{rgb}{0,0.7,0}
\newcommand\ra{\rangle}
\newcommand\la{\langle}
\newcommand\nn{\nonumber}
\newcommand\f{\frac}
\newcommand\p{\partial}

\usepackage{epsfig,amsopn}
\usepackage{graphicx}

\begin{document}

\title{Boltzmann entropy of a freely expanding quantum ideal gas}

\author{Saurav Pandey\textsuperscript{1}, Junaid Majeed Bhat\textsuperscript{1}, Abhishek Dhar\textsuperscript{1}, Sheldon Goldstein\textsuperscript{2},  David A. Huse\textsuperscript{3}, Manas Kulkarni\textsuperscript{1}, Anupam Kundu\textsuperscript{1} and Joel L.  Lebowitz\textsuperscript{2}}

\institute{${}^{1}$ International Centre for Theoretical Sciences, Tata Institute of Fundamental Research,
Bangalore 560089, India \\
	${}^{2}$  Departments of Mathematics and Physics, Rutgers University, Piscataway, NJ 08854, USA \\ ${}^{3}$  Physics Department, Princeton University, Princeton, NJ, 08544, USA}

\date{\today}
\authorrunning{Boltzmann entropy of a freely expanding quantum ideal gas}
\maketitle

\begin{abstract}
We study the time evolution of the Boltzmann entropy of a microstate during the non-equilibrium free expansion of a one-dimensional quantum ideal gas. 
This quantum Boltzmann entropy, $S_B$, essentially counts the ``number" of independent wavefunctions (microstates) giving rise to a specified macrostate.  It 
generally depends on the choice of macrovariables, such as the type and amount of coarse-graining, specifying a {\it non-}equilibrium macrostate of the system, but its extensive part agrees with the thermodynamic entropy in thermal equilibrium macrostates.  We examine two choices of macrovariables: 
the $U$-macrovariables are local observables in position space, while the $f$-macrovariables also include structure in momentum space.  For the quantum gas, we use a non-classical choice of the $f$-macrovariables. 
 For both choices, the corresponding entropies $s_B^f$ and $s_B^U$ grow and eventually saturate. As in the classical case, 
the growth rate of $s_B^f$ depends on the momentum coarse-graining scale.  If the gas is initially at equilibrium and is then released to expand to occupy twice the initial volume, the per-particle increase in the entropy for the $f$-macrostate, $\Delta s_B^f$, satisfies $\log{2}\leq\Delta s_B^f\leq 2\log{2}$ for fermions, and $0\leq\Delta s_B^f\leq\log{2}$ for bosons. For the same initial conditions, the change in the entropy $\Delta s_B^U$ for the $U$-macrostate is greater than $\Delta s_B^f$ when the gas is in the quantum regime where the final stationary state is not at thermal equilibrium.
\end{abstract}


\tableofcontents

\section{Introduction} \label{sec:intro}

Understanding the emergence of the second law for macroscopic systems from the reversible microscopic dynamics was  the remarkable achievement of Boltzmann~\cite{boltzmann1897,feynman2017character,lanford1976derivation,penrose1990emperor,greene2004fabric,lebowitz1993macroscopic,Lebowitz_PT1993,Goldstein_PD2004,griffiths1994}.  
He constructed an entropy function, satisfying the second law, for a typical individual microstate $X$ of a macroscopic system that is in a macrostate specified by the values of a collection of macrovariables.  This Boltzmann entropy 
is defined for systems both in and out of equilibrium 
as the logarithm of the ``number" of microstates corresponding to the system's macrostate.  For a classical system, the microstates are specified by points $X$ in phase space, and the Boltzmann entropy of $X$ is, up to an additive constant, the logarithm of 
the phase space volume of the set of all microstates for which the macrovariables have the same values as they do for $X$. For a quantum system, the microstates are specified by wave functions $|\Phi\rangle$.  This difference, which allows observables to have indeterminate values 
in the quantum microstate, can make the precise definition of quantum macrovariables and macrostates somewhat subtle~\cite{goldstein2017,tasaki2016,mori2018,de2006}, and also allows for new non-classical choices of macrovariables, as we illustrate below.  The Boltzmann entropy of $|\Phi\rangle$ is then the logarithm of the number of independent wave functions that have the same values of the {chosen} macrovariables as $|\Phi\rangle$ does.  At thermal equilibrium, {for all proper choices of macrovariables} this definition coincides, in its extensive part, with the thermodynamic entropy of Clausius~\cite{callen1998}.  

In a previous paper  some of the authors investigated the time evolution of the Boltzmann entropy for a freely expanding {\it classical} ideal gas, using two different choices of macrovariables~\cite{chakraborti2021entropy}.  In the present paper, we extend this to a freely expanding {\it quantum} ideal gas.  
We provide numerical results for the evolution of the microstates and the associated entropy functions during free expansion. We consider both fermions and bosons and present results in the low and high-temperature regimes. 
	
Before moving on, we 
mention some of the earlier works relevant to our study. Quantum quench of non-interacting fermions under various protocols have been studied using Wigner functions in \cite{dean2018fermions,manas2018}.  {Lattice fermions evolving from domain wall initial conditions have been studied with a focus on the evolution of the density profile and the growth of entanglement entropy~\cite{eisler2013,scopa2021}.} The diagonal entropy for {isolated quantum systems has been studied~\cite{rigol2011} for both integrable and non-integrable cases.}  {Some other recent relevant discussions of entropy in quantum systems   in the nonequilibrium setting can be {found in Refs.~\cite{deutsch2019a,deutsch2019b,solano2016theory}}.}

The rest of the paper is organized as follows. In Sec.~\eqref{sec:entropy} we discuss ideas related to the definition of macrovariables, macrostates and the Boltzmann entropy in quantum systems. 
In Sec.~\eqref{sec:example} we introduce the specific example that we will be analyzing in the paper. 
In Sec.~\eqref{sec:micro_evolve} we define our model of a quantum ideal gas and discuss aspects of its microscopic evolution. In particular, we discuss  properties of the single-particle density matrix in Sec.~\eqref{subsec:den_mat} and the Wigner distribution function in Sec.~\eqref{subsec:wig_den}. In Sec.~\eqref{sec:U_macro} we discuss the $U$-macrovariables and the corresponding Boltzmann entropy $S_B^U$. In Sec.~\eqref{sec:f_macro} we define the single-particle basis  of  states that are localized both in position and momentum space. This is then used to define the $f$-macrovariables and the  Boltzmann entropy $S_B^f$ for the quantum gas. In Sec.~\eqref{sec:numerics}, we present the numerical results. The results for fermions are presented in Sec.~\eqref{subsec:fermions} and for bosons in Sec.~\eqref{subsec:bosons}. We discuss and summarize our findings in Sec.~\eqref{sec:conclusion}. 


\section{Boltzmann entropy for quantum systems}
\label{sec:entropy}
We now discuss the construction of macrostates and the Boltzmann entropy for isolated macroscopic quantum systems. For a more detailed discussion see Refs.~\cite{goldstein2017,tasaki2016,mori2018}. Consider a quantum system with a Hilbert space $\mathcal{H}$, whose wave function lies in an energy shell $\mathcal{H}_E$ of width $\Delta E\ll E$. As macro observables, one option is to choose a set of commuting coarse-grained operators $\{\hat{M}_k\}, ~k=1,\ldots,J$, 
meaning {that the eigenvalues of each operator} are grouped into ``bins" and all eigenvalues  {{of that operator}} within each bin are set equal. The simultaneous diagonalization of {all the $\hat{M}_k$ operators then} provides a decomposition of the accessible Hilbert space into a sum of orthogonal subspaces $\mathcal{H_{\nu}}$, 
\begin{align}
\mathcal{H}_E = \bigoplus_\nu \mathcal{H}_\nu, 
\end{align}
where $\nu = (\nu_1,\ldots, \nu_J)$ defines a macrostate, and $\mathcal{H}_\nu$ is the joint eigenspace of the $\hat{M}_k$ with eigenvalues $\nu_k$.  The $\mathcal{H}_\nu$ will be referred to as macro-spaces, with   $|\mathcal{H}_\nu| = {\rm dim}~ \mathcal{H}_\nu$  the dimension of the corresponding macro-space. One then associates a Boltzmann entropy $S_\nu= k_B \ln |\mathcal{H}_\nu|$ to each of the macro-spaces.   Let us denote by $\hat{P}_\nu$ the projector onto the space $\mathcal{H}_\nu$. Any microstate, which is a pure state $|\Phi\ra$, is said to be in the macrostate $\nu$ if $|\Phi\rangle$ is almost in $\mathcal{H}_\nu$, i.e. if $\la \Phi| \hat{P}_\nu |\Phi \ra \approx 1$, and then its Boltzmann entropy is given by $S_B(|\Phi \ra)=S_{\nu}$. It is expected for many systems, including the one we study here, that for appropriate physical macrovariables if the initial pure state $|\Phi(0)\rangle$ is in a given macrostate, the  time-evolved microstate $|\Phi(t)\rangle$ will continue to be (at almost all times) in one single macrostate, i.e. 
Schrodinger cat-like states will not occur. The exception being when the system crosses from one macrostate to another.

 As in the classical case, for a physical choice of the macrovariables, a particular macro-space has by far the largest dimension, which we refer to  as the equilibrium macro-space and denote it by $\mathcal{H}_{\rm eq}$. It is characterized by the fact $|\mathcal{H}_{\rm eq}| = 
(1-\epsilon)|\mathcal{H}_E|, ~\epsilon \ll 1$, which we take as a physical requirement for any proper choice of the macrovariables. We also say that a system is in equilibrium when its microstate $|\Phi\rangle$  is in, or almost in, $\mathcal{H}_{\rm eq}$.  

Now consider the unitary time-evolution of a system that is initially prepared in a non-equilibrium pure quantum state $|\Phi(0)\ra$. 
Non-equilibrium  means that $|\Phi(0)\ra$ is not in, or {not} almost in, the space $\mathcal{H}_{\rm eq}$.  It starts in one of the other macro-spaces $\mathcal{H}_{\nu \neq {\rm eq}}$ and with time it moves between different macro-spaces until it eventually ends up in the equilibrium macro-space and stays there for almost all subsequent times.  It is expected that the non-equilibrium system should evolve to macro-spaces of higher dimensions, leading to a monotonic growth of entropy. This is what we would like to demonstrate in an explicit example. It is important to note that 
we are able to define this Boltzmann entropy for the pure quantum state, $|\Phi(t)\ra$, at any time~\cite{griffiths1994}.  

{\bf Computing the Boltzmann entropy}:~We note that the entropy $S_\nu$ of the macrostate $\nu$, and thus $S_B(|\Phi(t) \ra)$ {for $|\Phi(t) \ra$ in ${\cal{H}}_\nu$}, is equal to the Gibbs-von Neumann entropy of 
a generalized microcanonical ensemble, which is a mixed state that is uniform over (the unit sphere of) 
${\cal{H}}_\nu$. 
It is important to stress, however, that this generalized microcanonical ensemble is not {what we take as an} accurate microscopic description of the microstate; it is only being used as a construction to compute the Boltzmann entropy of the microstate {$|\Phi(t)\rangle$ at time $t$.} 
The Boltzmann entropy at a later time $t'> t$ is dictated by what macrostate the unitarily time-evolved microstate $|\Phi(t')\rangle$ at that time corresponds to. If we were to instead unitarily time-evolve the density matrix of this out-of-equilibrium generalized microcanonical ensemble, its Gibbs-von Neumann entropy would not change in contrast to the second law of thermodynamics.  
A typical out-of-equilibrium microstate 
will, as time advances, typically move to other macrostates of higher Boltzmann entropy, so the Boltzmann entropy thereby does obey the second law for almost all microstates. The macrostate does not time-evolve unitarily, due to it being constructed via a coarse-graining with suitably chosen macrovariables.

This generalized microcanonical ensemble that is used to calculate the Boltzmann entropy of the macrostate at time $t$ may  be replaced by an equilibrium ensemble for a fictitious system where constraints have been imposed on all macrovariables to have particular values $\nu$.  Then, since that is a macroscopic equilibrium system, the extensive part of its microcanonical entropy should be equal to the extensive part of the Gibbs-von Neumann entropy of an equivalent generalized canonical ensemble. This shows that we can arrive at a correct (to leading order) count of the number of independent microstates in ${\cal{H}}_\nu$ by calculating the Gibbs-von Neumann entropy of an equivalent generalized canonical ensemble.  This is convenient because in many cases, calculating properties is simpler in the canonical ensemble than in the microcanonical ensemble. 

If we accept, as is argued above, that the Boltzmann entropy of the microstate $|\Phi(t)\rangle$ of a macroscopic system 
is equal, to leading order in system size, to the Gibbs-von Neumann entropy of a properly chosen generalized canonical ensemble, this allows us to skip the step of defining the generalized microcanonical ensemble.  We use the expectation values, $ \la \hat{M}_k \ra=\la \Phi(t)|\hat{M}_k|\Phi(t) \ra, ~k=1,\ldots J,$ to define the ``equivalent" generalized canonical (GC) ensemble as
\begin{align}
\hat{\rho}_{\rm GC}&=\f{e^{-\sum_k \lambda_k  \hat{M}_k}}{Z_{\rm GC}}, \label{rhoGC} \\ {\rm where}~~ Z_{\rm GC}&= \Tr [e^{-\sum_k \lambda_k  \hat{M}_k}],
\end{align}
and the Lagrange multipliers $\lambda_k$ are obtained from the constraint equations
\begin{align}
 Tr[\hat{M}_k \hat{\rho}_{\rm GC}]=\la \hat{M}_k \ra, ~k=1,\ldots,J.
\end{align}
The resulting Boltzmann entropy $S_B$ for the microstate $|\Phi(t)\ra$ and the choice of $\{\hat{M}_k\}$ is then given by
\begin{align}
S_B= k_B \sum_k \lambda_k \la \hat{M}_k \ra + k_B \ln Z_{\rm GC}. \label{ent-GC}
\end{align}
We note that this is also equal to 
the maximum value of the Gibbs-von Neumann entropy 
\begin{align}
S_{\rm GvN}= -k_B \Tr [\hat{\rho}_N \ln \hat{\rho}_N]
\end{align}
over all $\hat{\rho}_N$, subject to the constraints $Tr[\hat{M}_k \hat{\rho}_{N}]=\la \hat{M}_k \ra$~\cite{fick1990}.

{Using this generalized canonical approach to {compute} the Boltzmann entropy permits more flexibility in the choice of macrovariables, as compared to those that adhere to the conditions needed for the above construction of generalized microcanonical macrostates.  In particular, the spectra of the operators $\{\hat{M}_k\}$ need not be coarse-grained, and for quantum systems, these operators need not commute {and may be microscopic operators, as we demonstrate in the example below.}}  

\begin{figure}
    \centering
    \includegraphics[scale=0.48]{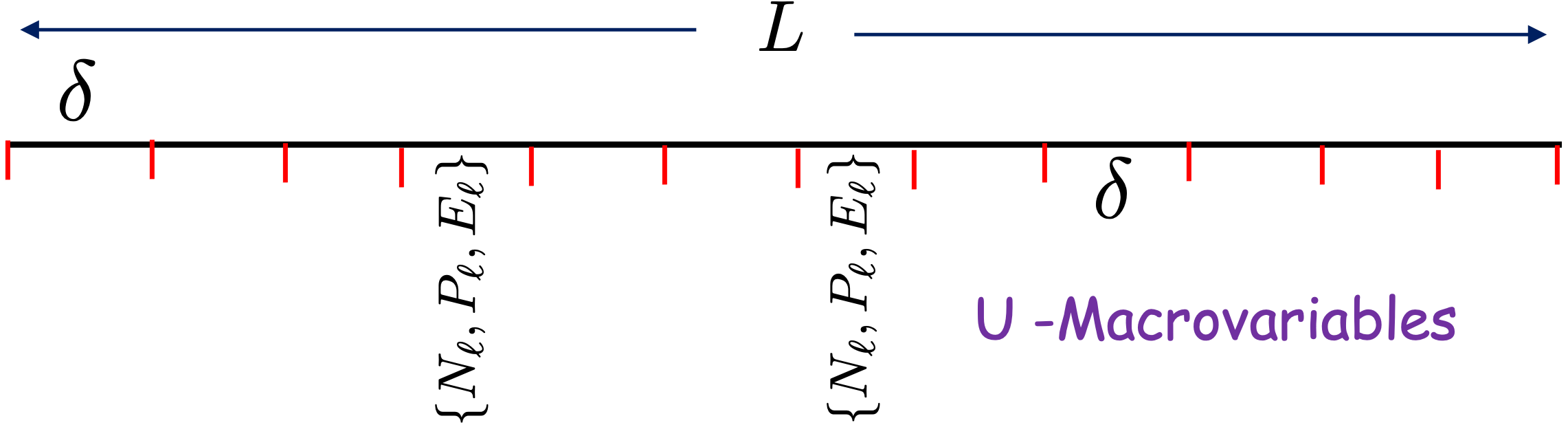}
    \includegraphics[scale=0.48]{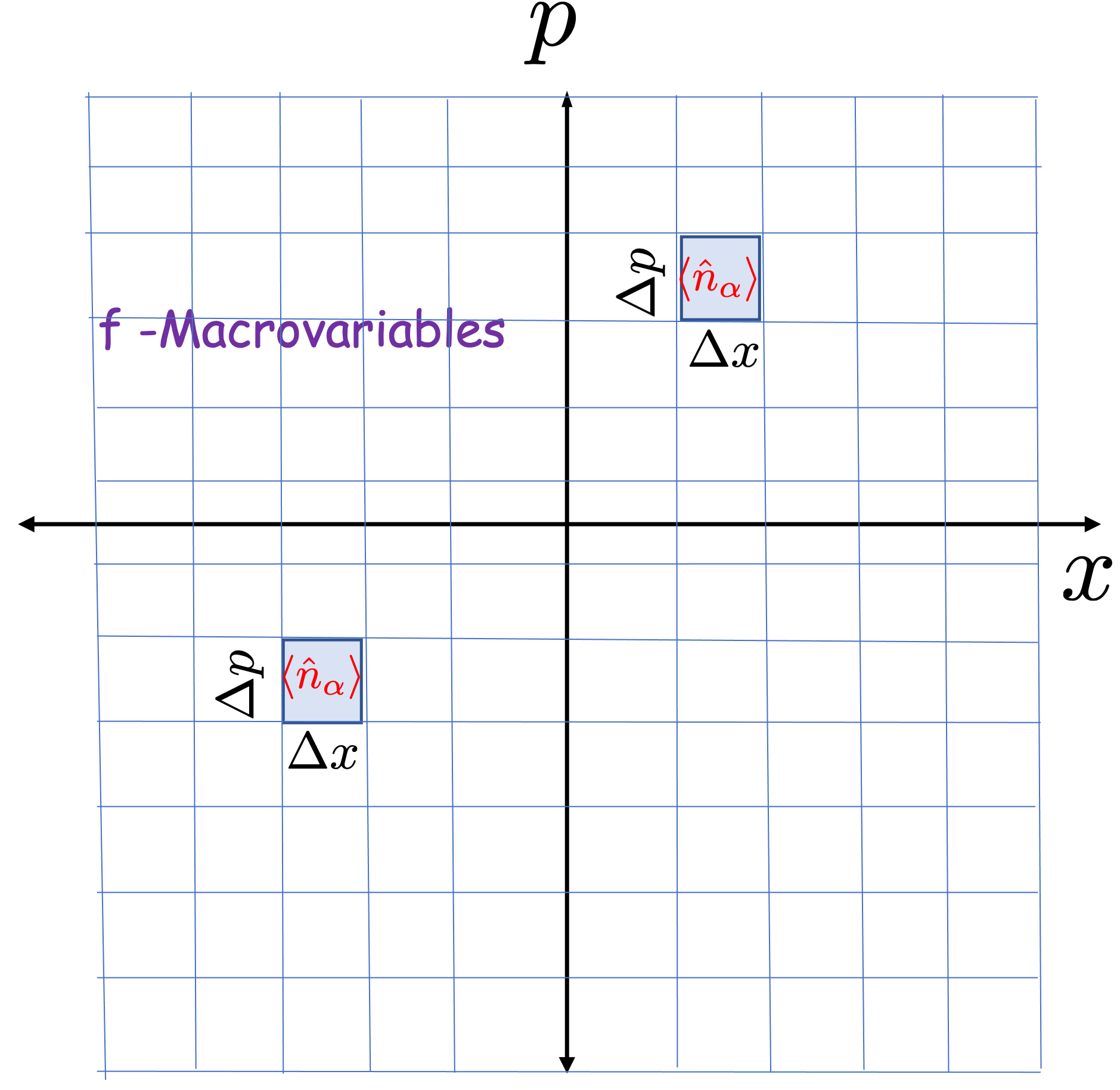}
    \caption{Schematic of the macrovariables that we consider. (Top panel) $U$-macrovariables in which, the box of length $L$  is divided into cells of length  $\delta$. By $U$-macrovariables, we mean coarse-grained particle number, momentum, and  energy within each cell. 
	(Bottom panel)
	$f$-macrovariables in which the macrovariables are defined by coarse-graining  in the single-particle phase space. By $f$-macrovariables we mean coarse-grained particle numbers within each box of area $\Delta x \Delta p$. A macrostate corresponds to a specification of the values of the macrovariables.}
    \label{fig:schematic}
\end{figure}

\section{An illustrative example: a one-dimensional quantum ideal gas}
\label{sec:example}
In this work, we consider a quantum ideal gas of $N$ particles on a circle of length $L$. {We examine the Boltzmann entropy for two different choices of macrovariables, schematically described in Fig.~\eqref{fig:schematic}, that are quantum analogs of those used for the classical ideal gas in Ref. \cite{chakraborti2021entropy}:} 
\begin{enumerate}
\item $U$-macrovariables: These are specified by the three conserved fields of particle number, momentum, and energy {and are direct analogs of the corresponding classical macrovariables \cite{chakraborti2021entropy}.}  They correspond to the usual fields that are used in the hydrodynamic description of interacting systems --- our non-interacting system has further conserved quantities but we can choose to  consider the coarse-grained description in terms of these fields alone. More specifically, we divide our system 
of length $L$ into $A$ spatial cells of size $\delta=L/A$. 
We consider a set of operators $\{\hat{N}_{\ell}, \hat{P}_{\ell}, \hat{E}_{\ell} \},~\ell=1,2\ldots A$, corresponding to 
the total number of particles, the total momentum, and the total energy in the $\ell^{\text{th}}$ cell. The precise definitions of these operators are given in Sec.~(\ref{sec:U_macro}).  Given that the system is in a microstate $|\Phi(t)\ra$, let $\la\hat{B}\ra=\la\Phi|\hat{B}|\Phi\ra$ denote  the expectation value of any operator $\hat{B}$. Then, the set of expectation values $\{N_{\ell},P_{\ell},E_{\ell}\}=\{\la \hat{N}_{\ell}\ra, \la\hat{P}_{\ell}\ra, \la\hat{E}_{\ell}\ra \}$ 
specify, as we argued in the preceding paragraphs, the $U$-macrostate. 

\item $f$-macrovariables:  {For the classical gas, the $f$-macrovariables are the distribution $f(x,p)$ of the single-particle positions and momenta~\cite{chakraborti2021entropy}.  Here we do not use a direct quantum analog of the classical macrovariables, but instead use a more microscopic choice, which is permitted for the quantum gas:}  
We choose a complete and orthonormal basis of single-particle wavefunctions, each of which is {approximately localized in position space within a width $\Delta x$ and in momentum space} with width $\Delta p$.  Such states will necessarily have to satisfy the uncertainty relation $\Delta x \Delta p \gtrsim \hbar$. As we will show in Sec.~\eqref{sec:f_macro}, it is possible to construct such an orthonormal basis set on the circle with states that are approximately localized in position space with $\Delta x= L/K$ and in momentum space with $\Delta p = 2 \pi \hbar K/L$, where $K$ is an integer. 
We denote these single-particle basis states by $|\psi_\alpha\ra$ with $\alpha\equiv (r,v)$, where $r = 1, \ldots, K$ and $v = \vartheta K$ where $\vartheta \in \mathbb{Z}$. These states are localized around $x = r \Delta x$ and $p = \vartheta \Delta p$. Let $\hat{n}_\alpha$ be the number operator corresponding to the occupation of the state $\ket{\psi_{\alpha}}$. The commuting operators $\{\hat{n}_\alpha\}$ form a set of {{\it micro}variables, since specifying all} of their eigenvalues specifies the microstate completely. 

To define macrovariables {for the quantum gas, we} consider the expectation values $\{ D_\alpha = \expval{\hat{n}_\alpha} \}$ in the microstate $\ket{\Phi}$. Since a pure state wavefunction of a quantum system implies a probabilistic description, specifying only the $\{ D_\alpha\}$ implies a 
coarse-graining, in the sense that they do not specify the microstate. 
{In fact, these expectation values $\{ D_\alpha\}$ constitute a good set of macrovariables that can be used to define a corresponding generalized canonical ensemble and Boltzmann entropy, following the procedure outlined above.  The resulting Boltzmann entropy is} 
\begin{align}
\hspace{0.6 cm} S_{\text{B}}^{f} = \sum_\alpha \left[-D_\alpha \ln D_\alpha \mp  (1\mp D_\alpha) \ln (1 \mp D_\alpha)\right],  \label{sBf-Dalpha}
\end{align}
where the negative and positive signs in the $\mp$ are for fermions and bosons, respectively.  {This entropy can also be described from a generalized {\it grand} canonical ensemble, where each of the single-particle states $\alpha$ has its own chemical potential $\mu_{\alpha}$ in order to set $D_{\alpha}$.}
{For the classical gas, the cells $\Delta x\Delta p$ used to define the $f$-macrovariables must be large enough so that they are typically occupied by many particles, in order to have enough coarse-graining~\cite{chakraborti2021entropy}.  For the quantum gas, on the other hand, fully microscopic cells may be used, and the needed coarse-graining is provided by using the expectation values as the macrovariables.} 

\end{enumerate}

\section{Microscopic model and dynamical evolution} \label{sec:micro_evolve}
Our system is  a quantum ideal gas of $N$ particles on a circle of length $L$.  The microstate of the system is a pure state $|\Phi(t)\rangle$ or, using the position representation, the wavefunction $\Phi(x_1,x_2,\ldots,x_N,t)$. This state evolves in time via the Schr\"odinger equation $i \hbar \p_t \Phi= \hat{H} \Phi$ with the free circle Hamiltonian $\hat{H}=-[\hbar^2/(2m)] \sum_{\ell=1}^N \p_{x_\ell}^2$. 
The gas is initially confined to a part of the circle, of length $a L$ with $0 <a <1$ (usually $a=1/2$), which we refer to as a ``box''. 
The initial $N$-particle state $\ket{\Phi(0)}$ may be taken to be an eigenstate of the occupations of the single-particle energy eigenstates in the box.  We also compare our results to the initial grand canonical mixed states in the box.  At $t = 0$, the box walls are removed at both ends and the gas is allowed to freely expand on the full circle. As the gas expands, we are interested in the time evolution of the macrovariables $U$ and $f$  and of the corresponding entropies. 

The single-particle energy eigenstates and eigenvalues on the circle are given by
\begin{align}
&\varphi_n(x) = \displaystyle{\frac{1}{\sqrt{L}}} e^{2 \pi i n x/L} \;\text{for} \;x \;\in \;[0, L), \label{circle ef} \\
&\epsilon_n = \frac{1}{2m} \left(\frac{2 \pi n \hbar}{L}\right)^2, \label{circle ev}
\end{align}  
where $n$ runs over all integers. These are also momentum eigenstates with eigenvalues $p_n = 2 \pi n \hbar/L$. A complete  $N$ particle basis  of Fock states is specified by the occupations $|\{N_n\}\rangle$ with $N_n \in \{0,1\}$ for fermions and $N_n \in \{0,1,\ldots, \infty\}$ for bosons, constrained so that $\sum_n N_n =N$.   We will find it useful sometimes to use the language of second quantization and so we define the particle creation and annihilation operators $\hat{\Psi}^\dagger(x), \hat{\Psi}(x)$ which create or annihilate, respectively, a particle at position $x$. For fermions, they satisfy the anti-commutation  relation $\{\hat{\Psi}(x), \hat{\Psi}^\dagger(x')\}=\delta(x-x')$, while for bosons they satisfy  the commutation relation $ [\hat{\Psi}(x), \hat{\Psi}^\dagger(x')]=\delta(x-x')$. We also define the creation and annihilation operators $\hat{b}^\dagger_n$, $\hat{b}_n$, corresponding to the single-particle energy eigenstate $\varphi_n$. These are related to the position operators as $\hat{\Psi}(x) = \sum_n \hat{b}_n \varphi_n(x)$. In the Heisenberg representation, these operators have simple time evolution for our noninteracting gas 
\begin{align}
\hat{b}_n(t)=e^{-i \epsilon_n t/\hbar} \hat{b}_n(0). \label{b-ev}
\end{align}

As we will see in the subsequent sections, for our purposes, it suffices to study the single-particle density matrix and the corresponding Wigner function. Therefore, in the following subsections, we discuss the evolution of the density matrix and the Wigner function. 

\subsection{Density matrix} \label{subsec:den_mat}
For our analysis of the dynamics, we will use 
the single-particle density operator $\hat\rho_1$.  In the Heisenberg picture, and in the basis of the single-particle energy eigenstates, $\hat\rho_1$ is a matrix of operators, with operator-valued matrix elements
\begin{align}
\hat{\rho}_{1,mn}(t) \equiv \hat\rho_1(p_m,p_n,t)=\hat{b}^\dagger_n(t) \hat{b}_m(t)~.
\end{align}
Note that this is normalized so $\sum_n \hat{\rho}_{1,nn}(t) = \hat{N}$ is the operator for the total particle number. In the position basis, the matrix elements of $\hat{\rho}_1(t)$ are given by  
\begin{align}
&{\rho}_1(x,x',t) \equiv \bra{x} \hat{\rho}_1(t) \ket{x'} = N \int dx_2 dx_3 \ldots dx_N \notag \\ 
&\Phi^*(x',x_2,x_3,\ldots x_N,t) \Phi(x,x_2,x_3,\ldots x_N,t) \notag \\ 
&= \Tr [ \hat\Psi^\dagger(x') \hat\Psi(x) \hat{\rho}_N(t)],
\end{align}
{where $\hat\rho_N (t)$ is the many-particle density matrix.} For {this} given many-particle state $\hat\rho_N$, the single-particle density matrix is the matrix of expectation values of the single-particle density operator, which in the single-particle momentum eigenbasis we write as
\begin{align}
&\tilde{\rho}_1(p_m, p_n, t) \equiv 
\Tr [ \hat{b}^\dagger_n(t) \hat{b}_m(t) \hat{\rho}_N] \notag \\ 
&\equiv \expval{\hat{b}^\dagger_n(t) \hat{b}_m(t)} =e^{-i (\epsilon_m-\epsilon_n)t/\hbar} \tilde{\rho}_1(p_m, p_n, 0). \label{rho-mn-t}
\end{align}
Thus we can compute the evolution of $\hat{\rho}_1$ once we know the initial value, $\tilde{\rho}_1(p_m, p_n, 0)$,  which we now evaluate.

For the description of the initial state where the gas is confined to a box of length $aL$, we will need the following single-particle ``box'' energy eigenspectrum
\begin{align}
&\chi_s(x) = \sqrt{\displaystyle{\frac{2}{a L}}} \sin \frac{s \pi x}{aL} \;\text{for} \;x \;\in \;[0, aL], \label{box ef} \\
&e_s = \frac{\pi^2 \hbar^2 s^2}{2 m a^2 L^2}, \label{box ev}
\end{align}
where $s$ runs over all positive integers. The box and circle states are related by the transformation
\begin{subequations}
\begin{align}
\chi_s(x) &= \sum\limits_{n=-\infty}^{\infty} V_{sn} \,\varphi_n(x), \label{chi-to-phi} \\
V_{sn} &= \int\limits_0^{aL} \chi_s(x) \varphi_n^{\star}(x) \,dx \label{V def} \\
 &= \frac{\sqrt{2 a} s}{\pi (4 a^2 n^2 - s^2)} [e^{-2  \pi i n a} \cos s \pi - 1]. \nn
\end{align}
\end{subequations}
Note that since $\{\chi_s(x), \,s = 1, \ldots, \infty\}$ do not form a complete set for states on the circle, $V$ is not an invertible matrix.

 {As mentioned before, we are interested in the evolution of an $N$ particle Fock state with energy $E$. For large $N$ and $E$, it is equivalent and more convenient to work with a grand-canonical distribution  with $\beta$ and $\mu$ chosen according to
 \begin{subequations}
\begin{align}
N &= \sum\limits_{s=1}^{\infty} f(e_s, \beta, \mu), \label{N constraint} \\
E &= \sum\limits_{s=1}^{\infty} f(e_s, \beta, \mu) e_s, \label{E constraint}
\end{align} 
\end{subequations}
where $f(e, \beta, \mu)= \displaystyle{\frac{1}{e^{\beta(e - \mu)} \pm 1}}$ is the Fermi (+) /Bose (-) function.
}

{
 In our numerical implementation, we use the following protocol.
We prepare the system in a  pure state with
the single-particle density matrix  given by
 \begin{equation}
\hat{\rho}_1^{\rm P}(0) = \sum\limits_{s=1}^{\infty} n_s \,|\chi_s \rangle \langle \chi_s|, \label{puresdm}
\end{equation} 
{where the superscript ``P" in Eq.~\eqref{puresdm} denotes pure state and the set $\{n_s\}$ } is chosen from the grand-canonical distribution 
\begin{align}
    P(\{ n_s \}) &= \frac{1}{Z} e^{-\beta \sum\limits_s (e_s - \mu) n_s}, \\
    Z &= \prod_s [1 \pm e^{-\beta (e_s - \mu)}]^{\pm 1},
\end{align}
subject to the constraints $\sum_s n_s =N$ and $\sum_s n_s e_s=E$.
Here $n_s \in \{0,1\}$ for fermions (+) and $n_s\in \{0,1,\ldots,\infty\}$ for bosons (-). In practice, these constraints are difficult to satisfy exactly. Hence we chose the set $\{n_s\}$ such that these constraints are satisfied within some desired tolerance. We compare our results for the pure initial state with the mixed (thermal) state described by the single-particle density matrix 
 \begin{equation}
\hat{\rho}_1^{\rm M}(0) = \sum\limits_{s=1}^{\infty} f(e_s, \beta, \mu) \,|\chi_s \rangle \langle \chi_s|, \label{initial single den mat}
\end{equation}}
where the superscript ``M" in Eq.~\eqref{initial single den mat} denotes mixed thermal state. {Using the box-to-circle transformation in Eq.~(\ref{chi-to-phi}), we can write the density matrix in Eq.~(\ref{initial single den mat}) in terms of the circle eigenfunctions to give
\begin{align}
\tilde{\rho}_1^{\rm M}(p_m, p_n, 0) = \sum\limits_{s=1}^{\infty} f(e_s,\beta,\mu)  V_{s m} V^{\star}_{s n}. \label{initial single den mat 2}
\end{align}
Using Eq.~\eqref{rho-mn-t}, the evolution of the density matrix in Eq.~\eqref{initial single den mat 2} is thus given by:
\begin{align}
\hat{\rho}_1^{\rm M}(t) = \sum\limits_{s=1}^{\infty} f(e_s,\beta,\mu) \hspace{-0.3 cm} \sum\limits_{m, n = -\infty}^{\infty} \hspace{-0.3 cm} V_{s m} V^{\star}_{s n} \,e^{-i(\epsilon_m - \epsilon_n) t/\hbar} \,|\varphi_m \rangle \langle \varphi_n|. \label{evolved single den mat}
\end{align}
The pure state density matrix $\hat{\rho}_1^{\rm P}$ has a similar representation with $f(e_s,\beta,\mu)$ replaced by $n_s$ in Eqs.~\eqref{initial single den mat 2} and \eqref{evolved single den mat}.
Note that the normalization condition $\Tr \hat{\rho}_1(t) =  N$ implies $(V V^{\dagger})_{ss} = 1$. The density matrix in Eq.~(\ref{evolved single den mat}) and the corresponding representation of $\hat{\rho}_1^{\rm P}(t)$ are used to calculate local densities corresponding to conserved quantities.}

\subsection{Wigner distribution function} \label{subsec:wig_den}
The Wigner distribution function (WDF) was introduced by  Wigner~\cite{wigner1932quantum,hillery1984} as a quantum analogue of the phase space distribution in classical systems. In 1-d, on the infinite line, the WDF, denoted by $w(x,p,t)$, is defined as 
\begin{equation}
w(x, p, t) = \frac{1}{2 \pi \hbar} \int\limits_{-\infty}^{\infty} dy \;\rho_1 \left( x + \frac{y}{2}, x- \frac{y}{2}, t \right) e^{i p y/\hbar}. \label{wigner def real line}
\end{equation}
We point out that the above transformation is one-to-one, thus the WDF and the density matrix contain the same information. For noninteracting systems in the absence of external potentials, $w(x, p, t)$ satisfies the simple equation
\begin{equation}
\partial_t w(x, p, t) + \frac{p}{m} \partial_x w(x, p, t) = 0, \label{wigner equation real line}
\end{equation}
which is identical to the evolution equation of the single-particle phase space density in classical non-interacting systems. The solution of  Eq~.(\ref{wigner equation real line}) is simply given by boosting the initial profile
\begin{equation}
w(x, p, t) = w\left( x- \frac{p t}{m}, p, 0 \right). \label{wigner evolution real line}
\end{equation}
In our model, the gas is confined to a circle of length $L$ instead of the infinite line. Therefore, we have $x \;\in \;[0, L)$ and the allowed momenta, $p_n = 2 \pi n \hbar/L$, with integer $n$, are discrete. A natural extension of the definition in Eq.~(\ref{wigner def real line}) to the case of circular coordinates would be to replace $p$ by $p_n$ and restrict the integral from $0$ to $L$~\cite{mukunda1978algebraic,mukunda1979wigner,berry1977}. However, this extension  does not satisfy Eq.~\eqref{wigner equation real line} with $p$ replaced by $p_n$. It can be shown~\cite{dhar2022} that a modified definition of the WDF on the circle, that  obeys Eq.~\eqref{wigner equation real line}, can be obtained.  For this one needs to define a new momentum variable: 
\begin{equation}
q_n = \frac{\pi n \hbar}{L} =  \frac{p_n}{2}, \label{q-def}
\end{equation}
that takes both integer as well as half-integer values. The modified WDF on the circle is thus defined as
\begin{align}
w(x,q_n,t) &= \frac{1}{2L} \int\limits_{-L}^L dy \;\rho_1 \left( x + \frac{y}{2}, x- \frac{y}{2}, t \right) e^{i q_n y/\hbar}, \label{wigner def real space} \\
&= \frac{1}{L} \sum\limits_{m = -\infty}^{\infty} \tilde{\rho}_1 \left( q_n + \frac{p_m}{2}, q_n - \frac{p_m}{2}, t \right) e^{i p_m x/\hbar}, \label{wigner def momentum space}
\end{align}
where, in Eq.~(\ref{wigner def momentum space}), the sum is over even values of $m$ if $q_n$ is an integer and odd values of $m$ is $q_n$ is a half-integer. The inverse transform is given by
\begin{equation}
\tilde{\rho}_1(p_m, p_n, t) = \int\limits_0^L dx \;w \left( x, \frac{p_m + p_n}{2}, t \right) e^{-i(p_m - p_n)x/\hbar}. \label{weyl transform}
\end{equation}

It is easy to check that the Wigner function in Eq.~\eqref{wigner def real space} satisfies Eq.~\eqref{wigner equation real line} on the circle with $p$ replaced by $q_n$ and thus has a solution of the form in Eq.~\eqref{wigner evolution real line} with  periodicity $L$. Using Eq.~(\ref{evolved single den mat}) in Eq.~\eqref{wigner def momentum space}, we get
\begin{align}
w(x, q_n, t) &= \sum\limits_{s=1}^{\infty}  f(e_s, \beta, \mu) \hspace{-0.25 cm} \sum\limits_{\ell, m = -\infty}^{\infty} \hspace{-0.25 cm} V_{s \ell} V^{\star}_{s m} \varphi_{\ell}(x) \varphi^{\star}_m(x) \nn \\
&\times e^{-i (\epsilon_{\ell} - \epsilon_m)t/\hbar} \delta(\ell+m-n), \label{wigner density 3}
\end{align} 
for the mixed state and a similar expression for the pure state with $f(e_s,\beta,\mu)$ replaced by $n_s$.
As we shall see in the next section, the WDF provides a simple way to define particle, momentum, and energy densities which are needed for defining the $U$-macrostate.

\section{Choices of macrostates and the corresponding entropies}
\subsection{$U$-macrostate and $S_B^U$} 
\label{sec:U_macro}
In this section, we present the details of the construction of the $U$-macrostate. In this description, the observables that define the system's macrostate are the expectation values of the three conserved macroscopic fields, namely the particle, momentum, and energy densities. We first motivate the basic definition of the operators corresponding to these observables and then show that their expectation values can be written in a simple and physically intuitive form in terms of the WDF.   We then discuss the corresponding entropy $S^U_B$.

\label{subsec:densities}
Using second quantized notation, the operators corresponding to the total number of particles, total momentum, and total energy on the circle,  in units of $\hbar = m = 1$, are given by
\begin{subequations}
\begin{align}
&\hat{N} = \int\limits_{0}^{L} dx\, {\hat{\Psi}^\dagger(x, t) \hat{\Psi}(x, t)}, \label{Ntotal} \\
&\hat{P} = -i\int\limits_{0}^{L} dx\, {\hat{\Psi}^\dagger(x, t) \partial_x \hat{\Psi}(x, t)}, \label{Ptotal} \\
&\hat{E} = -\frac{1}{2}\int\limits_{0}^{L} dx \,{\hat{\Psi}^\dagger(x, t)\partial_x^2 \hat{\Psi}(x, t)}. \label{Etotal}
\end{align}
\end{subequations}
It is then natural to define the following local density operators for the three fields:
\begin{subequations}
\begin{align}
\hat{n}(x,t) &= {\hat{\Psi}^\dagger(x, t) \hat{\Psi}(x, t)},  \label{particle density} \\
\hat{p}(x,t) &= \frac{i}{2} \bigg[ {(\partial_x \Psi^{\dagger}(x, t)) \Psi(x, t)} - {\Psi^\dagger(x, t)\partial_x \Psi(x, t)} \bigg], \label{P symmetrization} \\
\hat{e}(x,t) &= -\frac{1}{8}  \bigg[ \hat{\Psi}^{\dagger}(x, t) \partial_x^2 \hat{\Psi}(x, t) +  (\partial_x^2 \hat{\Psi}^{\dagger}(x, t)) \hat{\Psi}(x, t)\notag \\
&- 2 (\partial_x \hat{\Psi}^{\dagger}(x, t)) (\partial_x \hat{\Psi}(x, t)) \bigg]. \label{E symmetrization} 
\end{align}
\end{subequations} 
The forms above follow the requirement that the operators are self-adjoint, though, for the case of energy density, the choice is not unique. Our choice is motivated by the simple form it takes when we write the expectation value in terms of the WDF and it  satisfies conditions on the form of the density profile, that we expect on physical grounds. 
To define our coarse-grained macrovariables, we divide the circle into $A$ cells of size $\delta=L/A$ and label them by the index $\ell=1,2,\ldots, A$ with the $\ell^{\text{th}}$ cell beginning at $x=(\ell-1) \delta$. Our macrovariables are then the set of operators
\begin{align}
\{ \hat{N}_{\ell}, \hat{P}_{\ell}, \hat{E}_{\ell}\} = \int\limits_{(\ell-1) \delta}^{\ell \delta} dx \big{\{} \hat{n}(x), \hat{p}(x), \hat{e}(x)\big{\}}. \label{U-var}
\end{align}
To find the values of the macrovariables for a given microstate, we need the expectation values of the above operators. We find that the expectation values for the densities take the following simple forms when written in terms of the WDF: 
\begin{subequations}
\label{npe}
\begin{align} 
n(x, t) &= \sum\limits_{n=-\infty}^{\infty} w(x, q_n, t) \label{n-def},\\
p(x, t) &= \sum\limits_{n=-\infty}^{\infty} q_n w(x, q_n, t) \label{p-def}, \\
e(x, t) &= \sum\limits_{n=-\infty}^{\infty} (q_n^2/2) w(x, q_n, t) \label{e-def}.
\end{align} 
\end{subequations}
As is clear from the above equations, the densities take a physically intuitive form as the marginals of the Wigner function. The expectation values of the macrovariables in Eq.~\eqref{U-var} are denoted by $\{N_{\ell}, P_{\ell}, E_{\ell}\}$ and are readily obtained by integrating Eq.~\eqref{npe} across cells.

\paragraph*{Entropy of the $U$-macrostate:} \label{subsec:SBU}
 ~We now return to our  goal of defining the entropy for the $U$-macrostate. 
Given the set $\{N_{\ell}, P_{\ell}, E_{\ell}\}, ~\ell = 1, 2, \ldots, A$, we need to find the number of microstates consistent with a given specification for the values of the set. Assuming small correlations between cells, the number of microstates with these constraints is simply the product of the number of possible microstates in each cell (with the local constraints). Hence we get for the entropy
\begin{align}
S^U_B=\sum_{\ell=1}^A  S(N_{\ell},P_{\ell},E_{\ell}), 
\end{align} 
{where $S(N_{\ell},P_{\ell},E_{\ell})$ is  the equilibrium entropy of the $\ell^{\text{th}}$ cell of size $\delta = L/A$ with the specified values of the conserved quantities $N_{\ell},P_{\ell},E_{\ell}$.}

\subsection{$f$-macrostate and $S_B^f$} 
\label{sec:f_macro}

As discussed in Sec.~\eqref{sec:entropy} the construction of the $f$-macrostate requires us to find a basis set of single particle wavefunctions that are localized both in position and momentum space.  We now discuss its construction. The $\varphi$-basis, defined in Eq.~(\ref{circle ef}), consists of momentum eigenstates on the circle that are completely delocalized in position space.  To construct our localized basis, we superimpose $K$ number of successive $\varphi$-states labelled by a central momentum $p = 2 \pi \hbar v/L$, and generate $K$ new states
\begin{align}
\ket{\psi_{\alpha}} \equiv \ket{r, v} = \displaystyle{\frac{1}{\sqrt{K}}} \sum\limits_{n \;\epsilon \;\mathcal{R}_v}  \ket{\varphi_n} \,e^{-i n \frac{2 \pi r}{K}}, \label{psi basis}
\end{align}
where $\mathcal{R}_v = \left\{ v-(K-1)/2, \ldots, v+(K-1)/2 \right\}$ and $\alpha \equiv (r, v)$ is a collective index for the basis states. Note that $r = 1, \ldots, K$ and $v = \vartheta K$ {where $\vartheta$ takes all integer values}. The $K$ resulting $\ket{r, v}$ states for a given $v$ are localized around $rL/K, \;r = 1, \ldots, K$ in the position space and around $2 \pi \hbar v/L$ in the momentum space. For this reason, we shall refer to the $\psi$-basis in Eq.~(\ref{psi basis}) as the \emph{wavepacket basis}.

We now define the \emph{wavepacket density}, denoted by $D_\alpha(t) =\expval{\hat{n}_\alpha}\equiv D(r, v, t)$, as the diagonal of the single-particle density matrix $\hat{\rho}_1(t)$ in the wavepacket basis
\begin{align}
D_\alpha(t) \equiv D(r, v, t) = \bra{r, v} \hat{\rho}_1(t) \ket{r, v}. \label{wp density}
\end{align}	 
It turns out that one can write down the wavepacket density in terms of the Wigner function as 
\begin{equation}
D(r, v, t) = \frac{1}{K} \sum\limits_{\ell, m \in \mathcal{R}_v} \int\limits_0^L dx \;w \left( x, \frac{\ell + m}{2}, t \right) e^{-2 \pi i(\ell - m) z}, \label{wp wigner}
\end{equation}
where $z = x/L - r/K$. Making the variable transformation $q = (\ell + m)/2$ and $n = \ell - m$, we can rewrite Eq.~(\ref{wp wigner}) as
\begin{equation}
D(r, v, t) = \frac{1}{K} \sum\limits_{q = v_{-}}^{v_{+}} \int\limits_0^L dx \;w(x, q, t) \sum\limits_{n = 2(q - v_{+})}^{2(q - v_{-})} e^{-2 \pi i n z}, \label{wp wigner 2}
\end{equation}
where $v_{\pm} = v \pm (K-1)/2$. Summing over $n$ yields
\begin{align}
&D(r, v, t) = \frac{1}{K} \sum\limits_{q = v_{-}}^{v_{+}} \int\limits_0^L dx \;w(x, q, t) G_K(q-v, x), \label{wp wigner 3} \\ 
&G_K(q, x) = e^{-4 \pi i q z } \,\frac{\sin (2K-1) \pi z}{\sin \pi z}, \label{G kernel} 
\end{align}
where recall $z = x/L - r/K$.
Eq.~(\ref{wp wigner 3}) suggests that the wavepacket density is nothing but a coarse-grained WDF. It is also instructive to   look at the marginals of the wavepacket density. The momentum marginal is given by the expression~(see Appendix~\ref{app:marginals})
\begin{align}
D_v(v) = \sum\limits_{r=1}^K D(r, v,t) = \sum_{n \in \mathcal{R}_v } \tilde{\rho}_1(p_n, p_n,0), \label{p marginal}
\end{align}
where recall $\mathcal{R}_v $ denotes set of momenta centered around $v$.
Similarly, the position marginal is given by~
(see Appendix~\ref{app:marginals})
\begin{align}
D_r(r,t) &= \sum\limits_v D(r, v,t)  = \int\limits_{0}^{L} dx \,h_K\left( \frac{x}{L} - \frac{r}{K} \right) \rho_1(x,x,t), \label{r marginal} \\
\text{with}~&h_K(x) = \displaystyle{\frac{1}{K^2}} \left( \frac{\sin \pi K z}{\sin \pi z} \right)^2. \label{h kernel}
\end{align}
 Note that Eqs.~(\ref{p marginal}, \ref{r marginal}) represent coarse-grained versions of the diagonal elements of the density matrix in the momentum and position basis.

\paragraph*{Entropy of the $f$-macrostate:}~We now discuss the entropy formula in Eq.~\eqref{sBf-Dalpha} for a given specification of the set $\{D_\alpha\}$. 
To derive this formula, one maximizes the Gibbs-von Neumann entropy, $S_{\rm GvN} = - \Tr [\hat{\rho}_N \log \hat{\rho}_N]$, subject to the constraint
\begin{equation}
D_{\alpha}(t) = \bra{\psi_{\alpha}} \hat{\rho}_1(t) \ket{\psi_{\alpha}} = \Tr [\hat{\rho}_N(t) \hat{\Psi}_{\alpha}^{\dagger} \hat{\Psi}_{\alpha}], \label{wp density constraint}
\end{equation}
where $\hat{\Psi}_{\alpha}$ creates a particle in the state $\ket{\psi_{\alpha}}$. 
The density matrix $\hat{\rho}_N^{\star}$ that maximizes $S_{\rm GvN}$ is given by 
\begin{align}
&\hat{\rho}_N^{\star} = \frac{1}{Z} \,e^{-\sum\limits_{\alpha} \lambda_{\alpha} \hat{\Psi}_{\alpha}^{\dagger} \hat{\Psi}_{\alpha}}, \label{maximal wp density matrix} \\
&\quad Z = \prod\limits_{\alpha} \,[1 \pm e^{-\lambda_{\alpha}}]^{\pm1}, \label{wp partition function}
\end{align} 
where $+$ is for fermions and $-$ is for bosons. Note that this is precisely the density matrix that defines the equivalent generalized canonical ensemble $\hat{\rho}_{GC}$, as mentioned in Sec.~(\ref{sec:entropy}). The Lagrange multipliers $\lambda_{\alpha}$ are given by the relation 
\begin{equation}
D_{\alpha} = \Tr [\hat{\rho}_N^{\star} \hat{\Psi}_{\alpha}^{\dagger} \hat{\Psi}_{\alpha}] = \displaystyle{\frac{1}{e^{\lambda_{\alpha}} \pm 1}}. \label{lagrange multipliers}
\end{equation}
We can thus write down the maximal Gibbs-von Neumann entropy, $S_B^f = - \Tr [\hat{\rho}_N^{\star} \log \hat{\rho}_N^{\star}]$, in terms of $D_{\alpha}$ as
\begin{equation}
S_B^f = \sum\limits_{\alpha} \left[ -D_\alpha  \ln D_\alpha \mp  (1\mp D_\alpha)  \ln (1\mp D_\alpha) \right], \label{wp entropy}
\end{equation}
where the $-$ in the $\mp$ is for fermions and the $+$ is for bosons.

In order to compute the final change in entropy  $\Delta s^f_B=(S_B^f(\infty) - S_B^f(0))/N$, we next provide analytical estimates for the values of $D_\alpha(t)$ at $t \to \infty$ and $t=0$. 
Substituting the expression for $\hat{\rho}_1(t)$ from Eq.~(\ref{evolved single den mat}) into Eq.~(\ref{wp density}), we obtain an explicit expression for the wavepacket density
\begin{align}
D_{\alpha}(t) = \sum\limits_{s=1}^{\infty} f(e_s, \beta, \mu) \hspace{-0.2 cm} &\sum\limits_{m, n = -\infty}^{\infty} \hspace{-0.2 cm} V_{s m} V^{\star}_{s n} e^{-i (\epsilon_m - \epsilon_n)t/\hbar}  \langle \psi_{\alpha} | \varphi_m \rangle \langle \varphi_n | \psi_{\alpha} \rangle. \label{wp density 2}
\end{align}
We first discuss the late time limit. As $t \to \infty $, we only get contributions from the $m = n$ terms in Eq.~(\ref{wp density 2})
\begin{equation}
D_{\alpha}(\infty) = \sum\limits_{s=1}^{\infty} f(e_s, \beta, \mu) \sum\limits_{n=-\infty}^{\infty} |V_{s n}|^2 \,||\langle \varphi_n | \psi_{\alpha} \rangle||^2. \label{wp density inf}
\end{equation}
In the limit of large $N$ and $L$ (keeping $N/L = \rho$ fixed), we use the explicit form of $V_{sn}$ in Eq.~\eqref{V def} and find that   $|V_{sn}|^2 $ is highly peaked around $s \approx 2 a  n$ for large $n$. Hence the sum can be approximated as 
$\sum_s f(e_s, \beta, \mu) \,|V_{s n}|^2 \approx f(e_{2 a n}, \beta, \mu) \sum_s  \,|V_{s n}|^2 = a f(\epsilon_n, \beta, \mu)$.
We thus get for the late time 
\begin{equation}
D_{\alpha}(\infty) = \frac{a}{K} \sum\limits_{n = \mathcal{R_v}} f(\epsilon_n, \beta, \mu). \label{wp density inf 2} 
\end{equation}
To compute  $D_\alpha(0)$ we start with the expression 
\begin{align}
D_{\alpha}(0) = \sum\limits_{s=1}^{\infty} f(e_s, \beta, \mu) \hspace{-0.2 cm} &\sum\limits_{m, n = -\infty}^{\infty} \hspace{-0.2 cm} V_{s m} V^{\star}_{s n}   \langle \psi_{\alpha} | \varphi_m \rangle \langle \varphi_n | \psi_{\alpha} \rangle. \label{wp density 2_to}
\end{align}
Using Eq.~\eqref{V def} for $V_{sm}$, we obtain the following simplified form:
\begin{align}
&D_{\alpha}(0) = \frac{1}{K}\sum\limits_{s=1}^{\infty} f(e_s, \beta, \mu)\abs{g_{v}(s,r)}^2,~\text{where}\label{wp density 3}
\\
&g_{v}(s,r) = \sqrt{\frac{2}{a}} \int_{-r/K}^{a-r/K} \hspace{-0.3 cm} dz \sin \left[ \frac{s \pi}{a} \left( z+\frac{r}{K} \right) \right] e^{-2\pi i v z} \frac{\sin{\pi K z}}{\sin{\pi z}}.
\end{align}
Since the integrand in $g_v(s,r)$ is highly peaked about $z=0$ for large $K$,  we make the replacement $\sin(\pi z) \approx \pi z$ in the numerator whenever $z=0$ falls inside the integration limit {\it i.e.} for $r<Ka$. This allows us to make the following approximation $g_v(s,r) \approx \Theta(Ka-r) \tilde{g}_v(s,r)$ where $\Theta(r)$ is the Heaviside Theta function and $\tilde{g}_v(s, r)$ is given by
\begin{align}
\tilde g_{v}(s,r) &\approx 
\sqrt{\frac{2}{a}}\int_{-\infty}^{\infty} dz \sin \left[ \frac{s \pi}{a} \left( z+\frac{r}{K} \right) \right] e^{-2\pi i v z} \frac{\sin{\pi K z}}{\pi z}.
\end{align}
This finally implies
\begin{align}
|\tilde g_{v}(s,r)|^2 \approx \frac{1}{2a}\Theta(s_+-s) \Theta(s-s_{-}) ,
\end{align}
with $s_\pm=2av\pm Ka$. Using 
this approximation in Eq.~\eqref{wp density 3}, we get
\begin{equation}
    D_{\alpha}(0) \approx
        \displaystyle{\frac{1}{2aK}} \sum\limits_{s = s_{-}}^{s_{+}} f(e_s, \beta, \mu)\Theta(r<Ka). \label{wp density initial}
\end{equation} 
Note that while making the above approximations, we have ignored possible $r$ dependence near the edges of the initial box of size $aL$. As a result, the approximated density $D_\alpha(0)$ slightly underestimates the initial entropy.

Using $D_\alpha(\infty)$ from Eq.~\eqref{wp density inf 2} and $D_\alpha(0)$ from Eq.~\eqref{wp density initial}, in Eq.~\eqref{wp entropy} one can numerically compute $\Delta s^f_B$ for different $T$ and $\mu$. {However the above procedure does not remain valid for bosons at very low temperatures because the particles occupy only a few low-lying energy states and consequently they do not relax. On the other hand for fermions, the above procedure works for all $T$, and specifically at $T=0$, one can obtain an explicit expression of $\Delta s^f_B$.  To obtain this expression we first note that for $T=0$, Eq.~(\ref{wp density inf 2}) gives $D_{\alpha}(\infty) = a$ for all $\alpha$ except those which are very close to the Fermi surface.}
Using $D_\alpha(\infty) \approx a$ in Eq.~\eqref{wp entropy}, we find 
\begin{equation}
S_B^f(\infty) = -\sum\limits_{\alpha} [a \ln a + (1 - a) \ln (1 - a)], \label{wp entropy inf}
\end{equation}
where the summation extends over $2 n_{max}$ terms such that $\epsilon_{n_{max}} = \mu$. Given that $e_{s} = \epsilon_{s/2a}$ and $e_{N} = \mu$, we get $n_{max} = N/2a$. The number of terms in the $\alpha$-sum is thus $N/a$ and hence, the late time entropy per particle is given by
\begin{equation}
s_B^f(\infty) = -\frac{1}{a} [a \ln a + (1 - a) \ln (1 - a)]. \label{wp entropy inf 2} 
\end{equation} 
For $a = 1/2$, we obtain $s_B^f(\infty) = 2 \ln 2$. The initial density in Eq.~(\ref{wp density initial}) simplifies in a similar manner and we get
\begin{equation}
D_{\alpha}(0) = \frac{1}{2aK} \sum\limits_{s = s_{-}}^{s_{+}} \Theta(e_s < \mu) \approx 1, \label{wp density initial 2}
\end{equation}
which results in the initial entropy at $T=0$ being zero (neglecting edge contributions as mentioned earlier). We therefore get $\Delta s_B^f (T = 0) = 2 \ln 2$ for the Fermi gas expanding to twice the initial volume.

\begin{figure*}
\centering

\subfigure[Fermi function]{
	\includegraphics[scale=0.4]{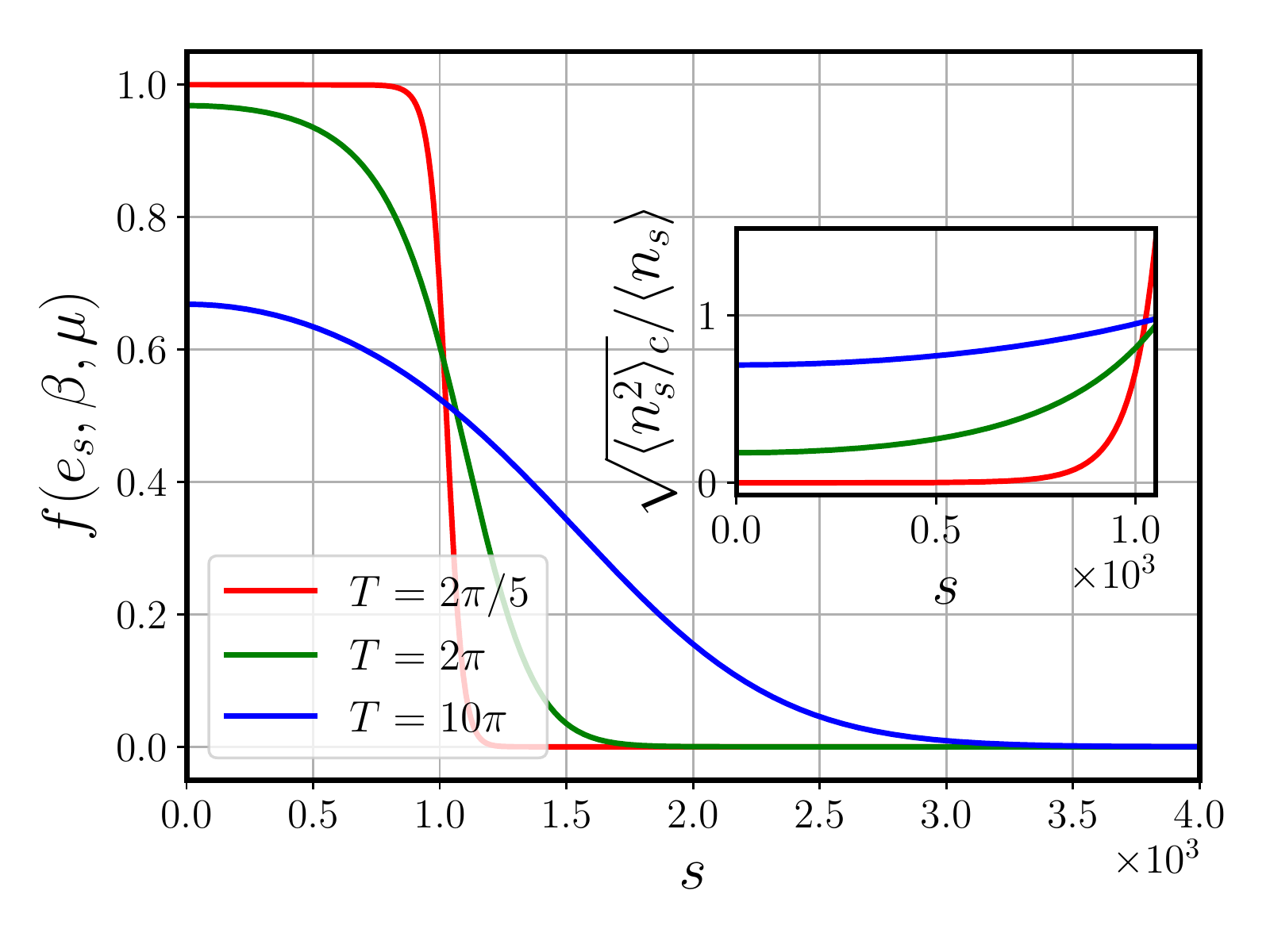}
}
\subfigure[Bose function]{
	\includegraphics[scale=0.4]{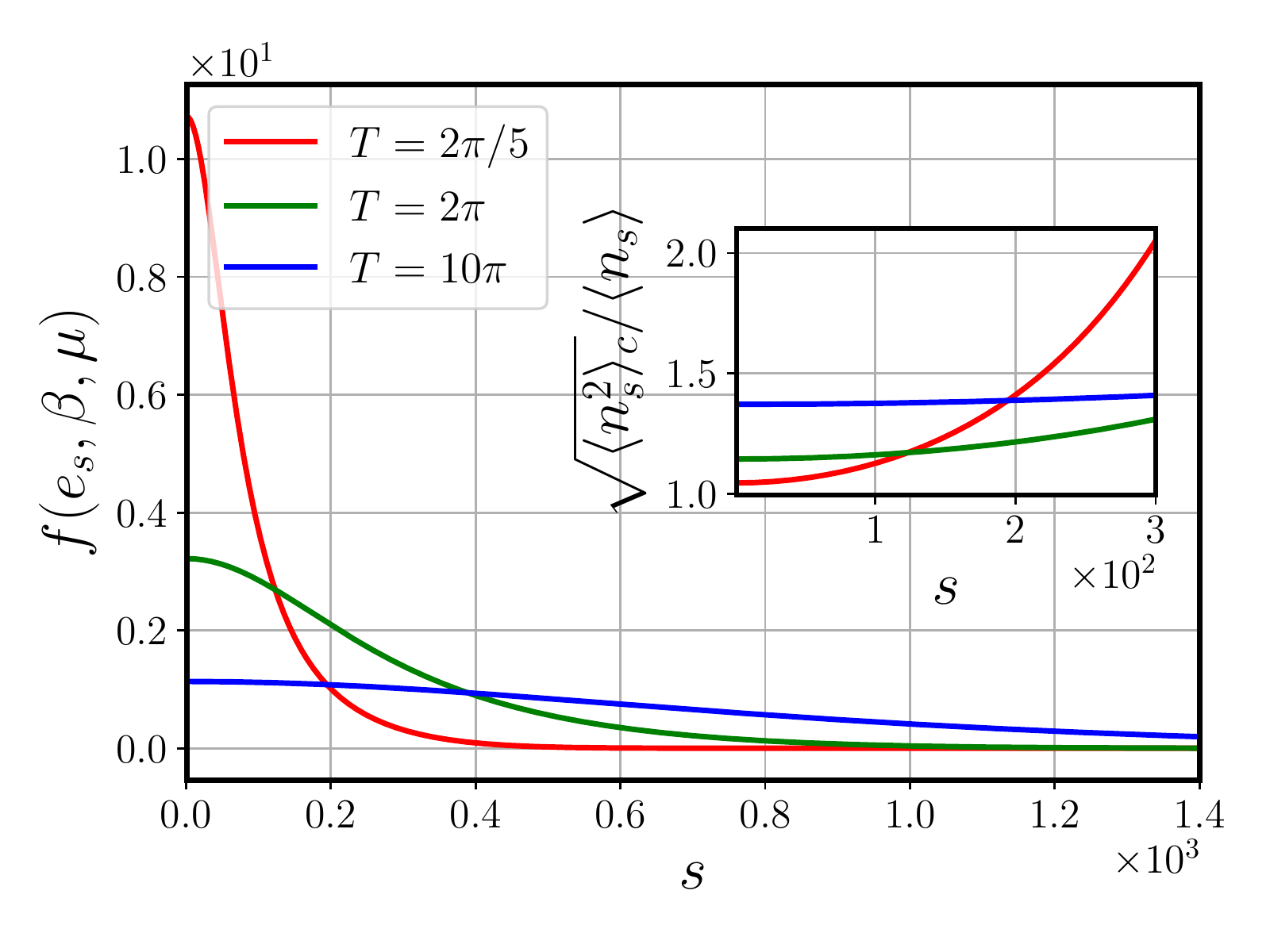}
}
\caption{Fermi and Bose distribution functions showing the mean occupation numbers of energy levels at the low and high temperatures (along with an intermediate temperature for comparison)  used in our numerical study. The particle density was set to $\rho=1$. The insets show the relative number fluctuations.} \label{fermi-bose}
\end{figure*}

\begin{figure*}
\centering

\subfigure[$T=2\pi/5$,~$\mu=19.84$]{\includegraphics[scale=0.3]{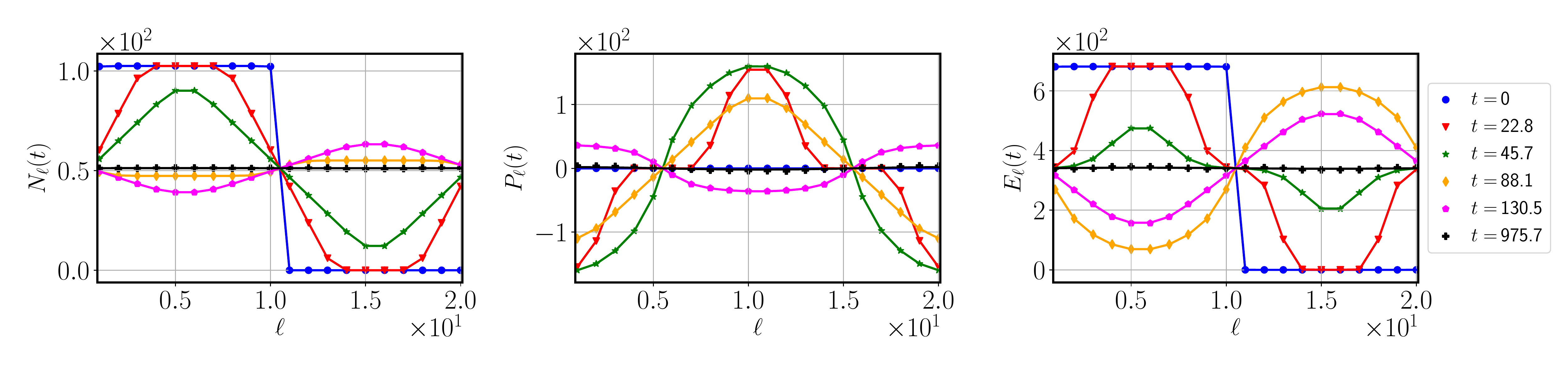}
}

\subfigure[$T=10\pi$,~$\mu=22.04$]{
	\includegraphics[scale=0.3]{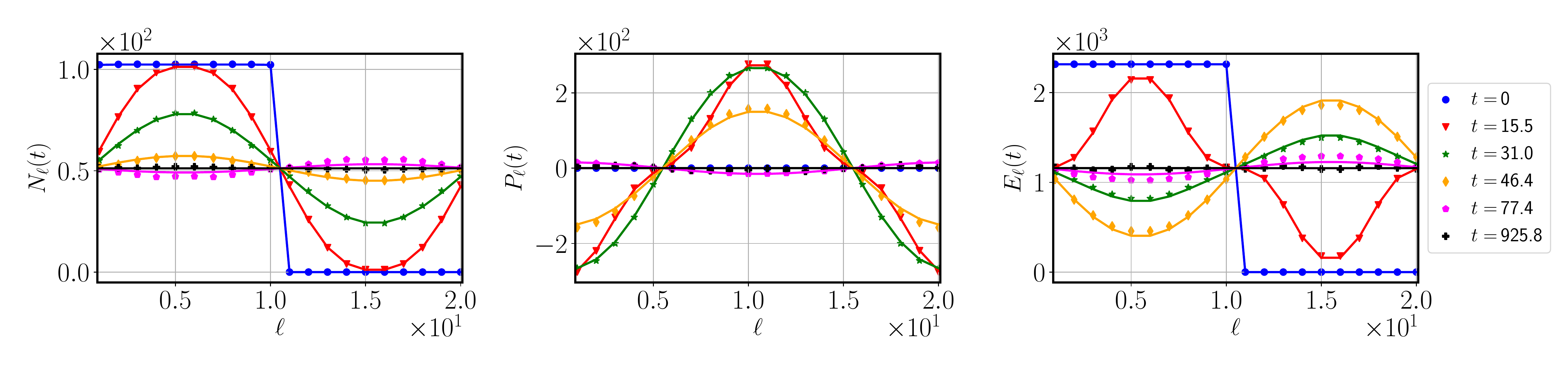}
}
\caption{{\bf Fermions - evolution of $U$-macrovariables}: Plots  showing the spatial profiles of the 
 number of particles, the total momentum, and the total energy  in each of the $A = 20$ cells for $N=L=1024$ at different times. The dots represent results for  the pure state initial condition while the solid lines correspond to a  thermal mixed state. Results are presented for  two different  temperatures $T=2 \pi/5,~10 \pi$ {(top and bottom row respectively)}, and chemical potentials are fixed so that the mean density is set at $\rho=1$. We see a good agreement between the pure state and the thermal state results. \label{2NPEfvsx}}
\end{figure*}

\begin{figure*}[t]
\subfigure[$T=2\pi/5$,~$\mu=19.83$]{
	\includegraphics[scale=0.43]{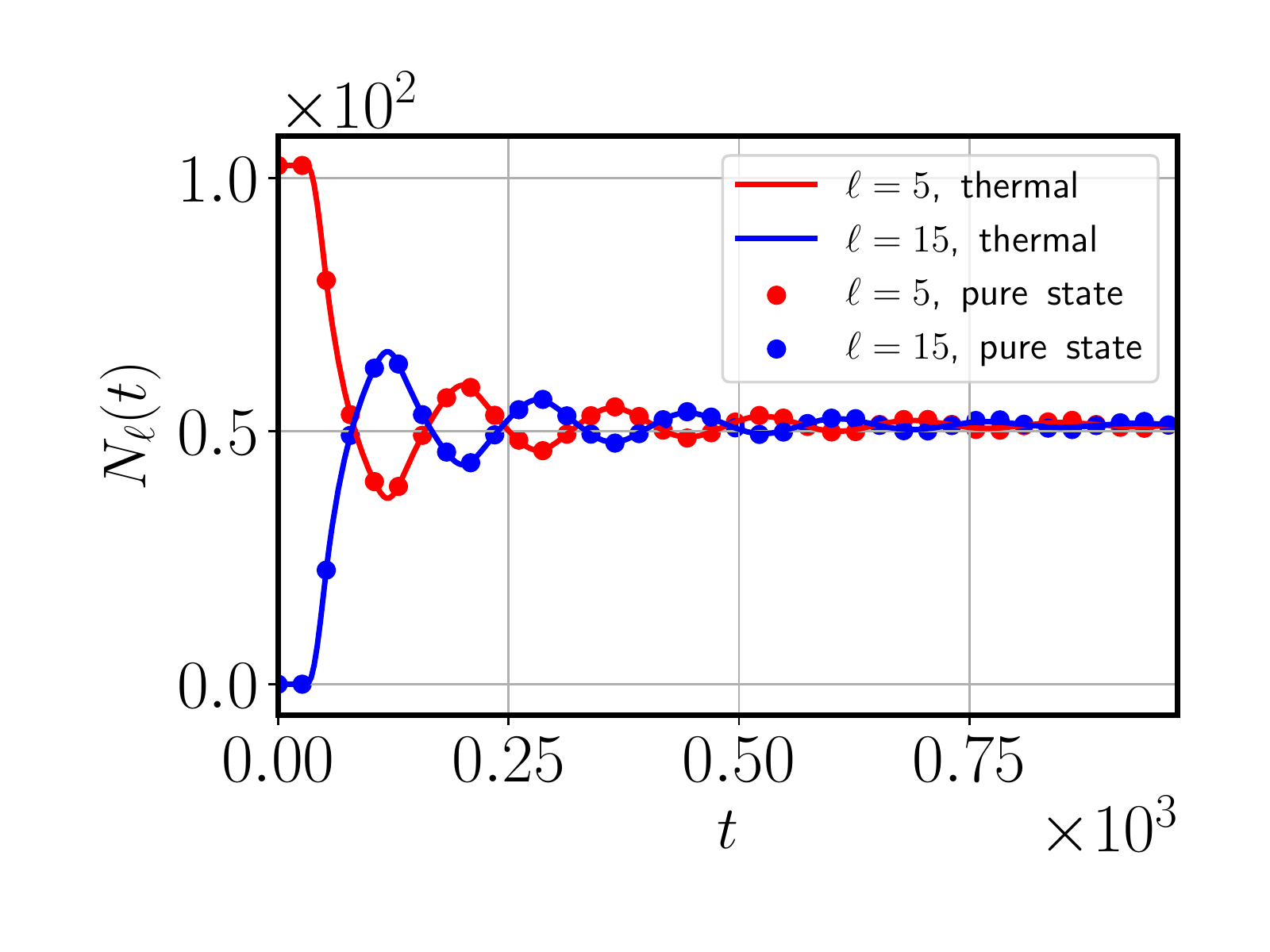}
}
\subfigure[$T=10\pi$,~$\mu=22.02$]{
	\includegraphics[scale=0.43]{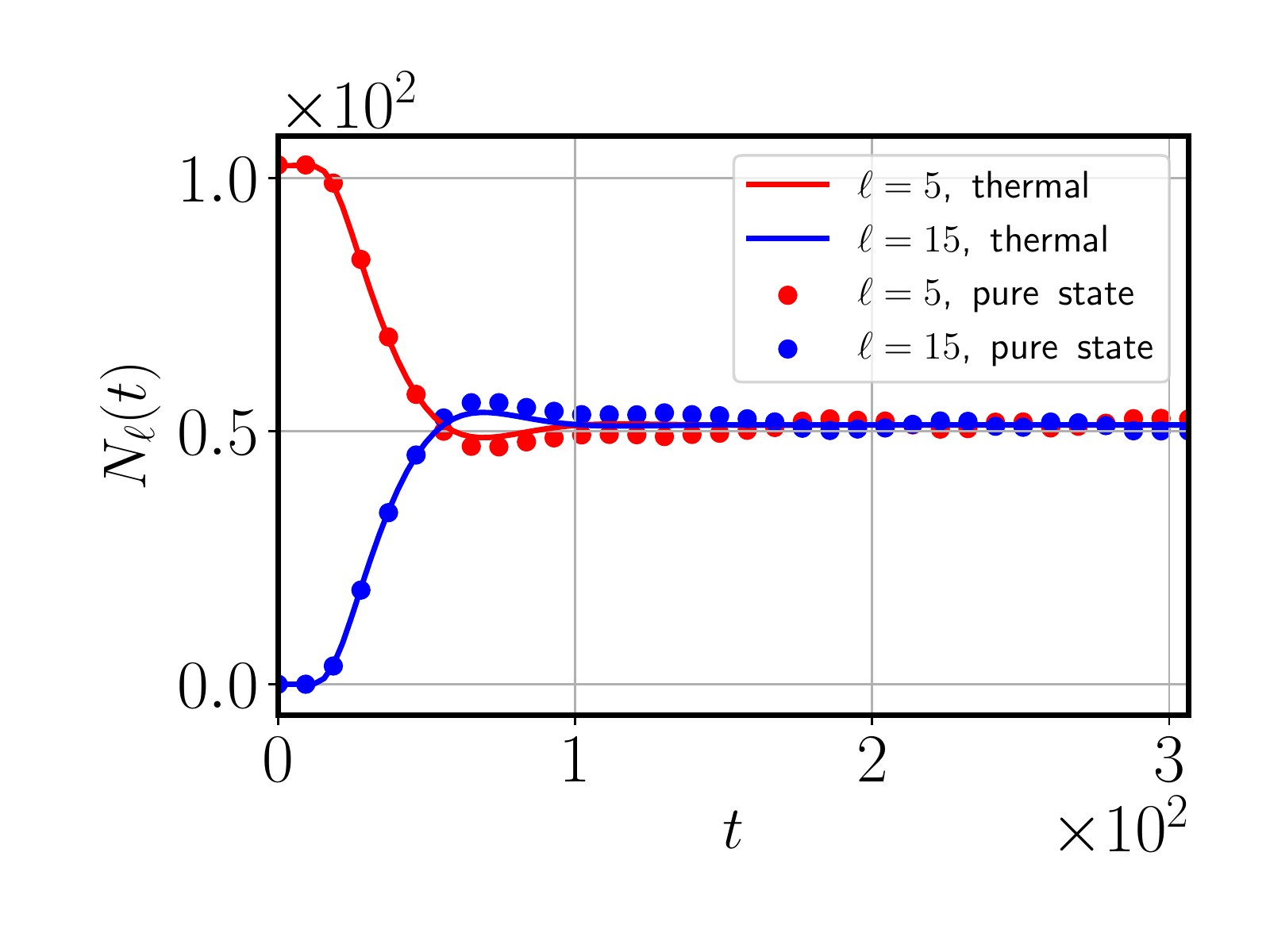}
}
\caption{{\bf Fermions}: Time evolution of the number of particles inside the $5^{th}$ and $15^{th}$ cells out of a total of $A = 20$ cells for $N = L = 1024$.  As in Fig.~\eqref{2NPEfvsx},  we see a good agreement between the pure state (dots) and the thermal state results (solid line). We see marked oscillations at low temperatures with a time period $\tau_p=L/v_f$ (see text). \label{fNvst}}
\end{figure*}

\begin{figure*}[t]
\subfigure[$T=2\pi/5$,~$\mu=19.83$,~$\Delta s = 1.46$]{
	\includegraphics[scale=0.43]{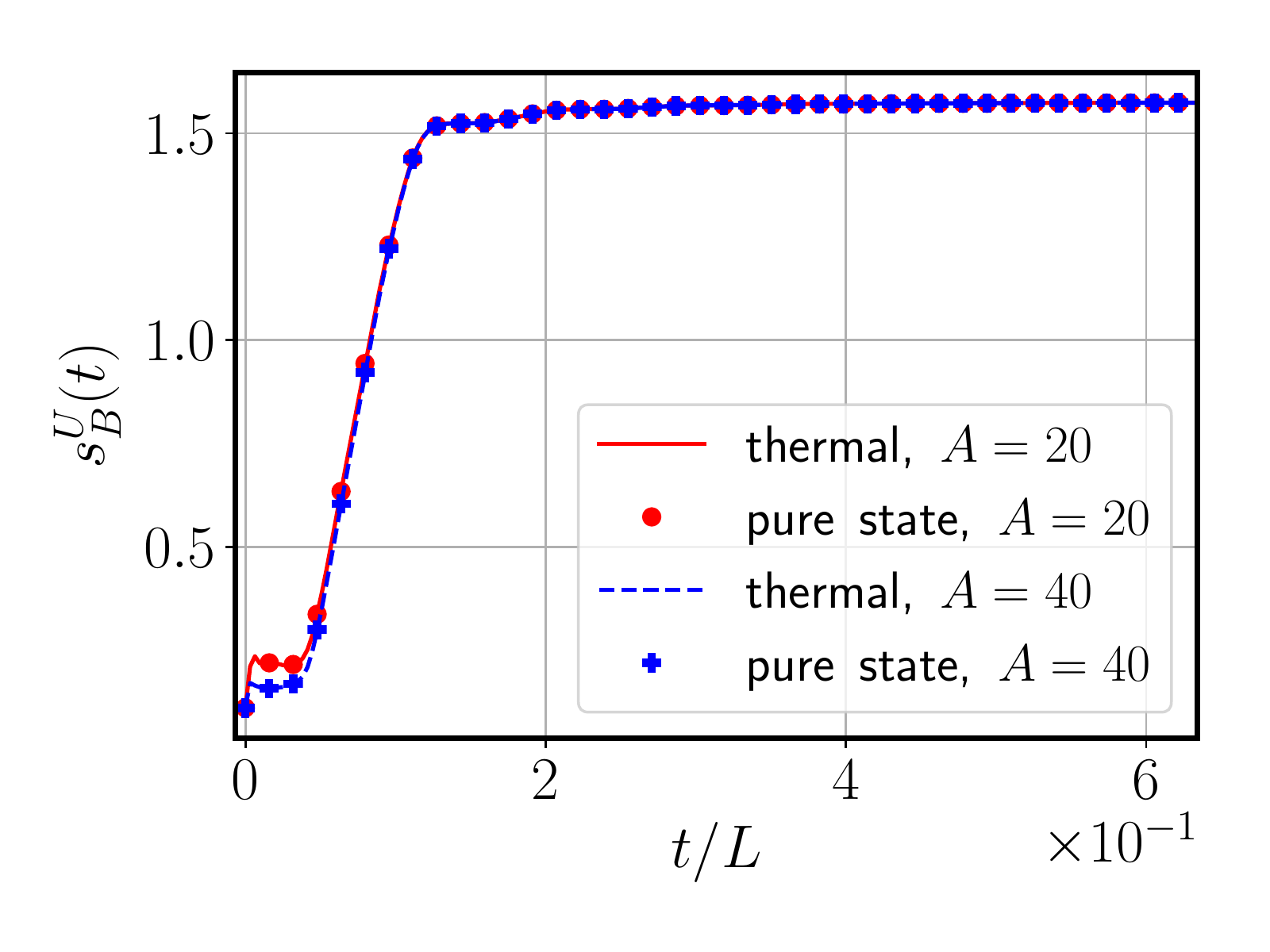}
}
\subfigure[$T=10\pi$,~$\mu=22.02$,~$\Delta s = 0.88$]{
	\includegraphics[scale=0.43]{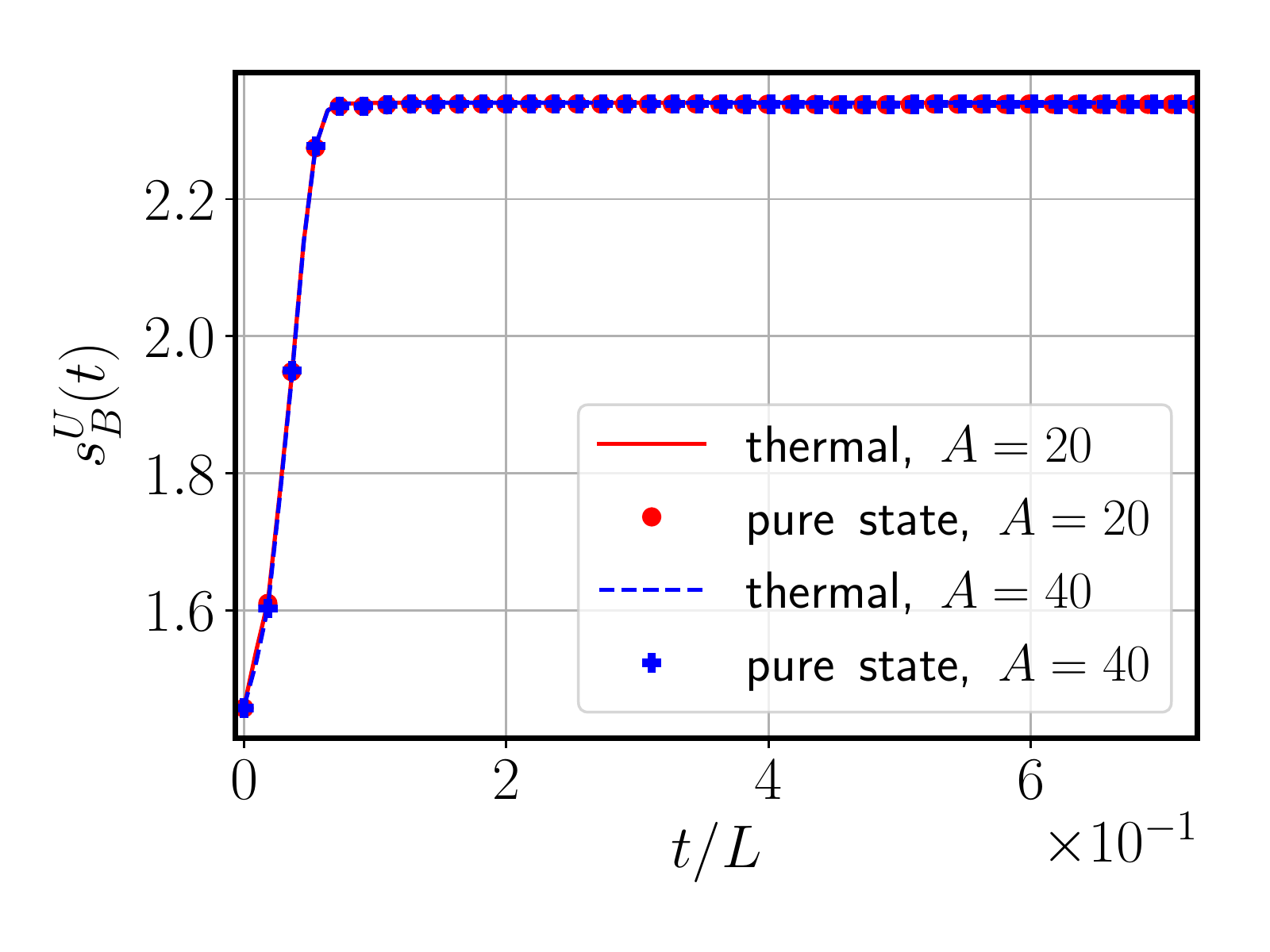}
}
\caption{{\bf Fermions}: $U$-macrostate entropy growth for fermions for $N = L = 1024$ at two different temperatures and for two different coarse-graining scales with $A=20$ (blue lines) and $A=40 $ (red lines). Results are presented for the pure state (dots) and  thermal state (solid lines) initial conditions. The initial sharp rise corresponds to the filling of the two empty cells on either side of the unfilled  part of the circle. This jump is smaller for finer coarse-graining size.  We also observe an initial flat profile in (a) which persists  till time $(L/4)/v_{\rm f}$, where $v_{\rm f}$ is the Fermi velocity. At large times, in all cases, the entropy saturates to the thermodynamic entropy of the new equilibrium state (corresponding to uniform values of the three conserved fields on the circle). The sub-captions give the values for the change in entropy per particle, $\Delta s$, for the two cases.} \label{SBUfplots}
\end{figure*}

\begin{figure*}
\subfigure{
	\includegraphics[scale=0.45]{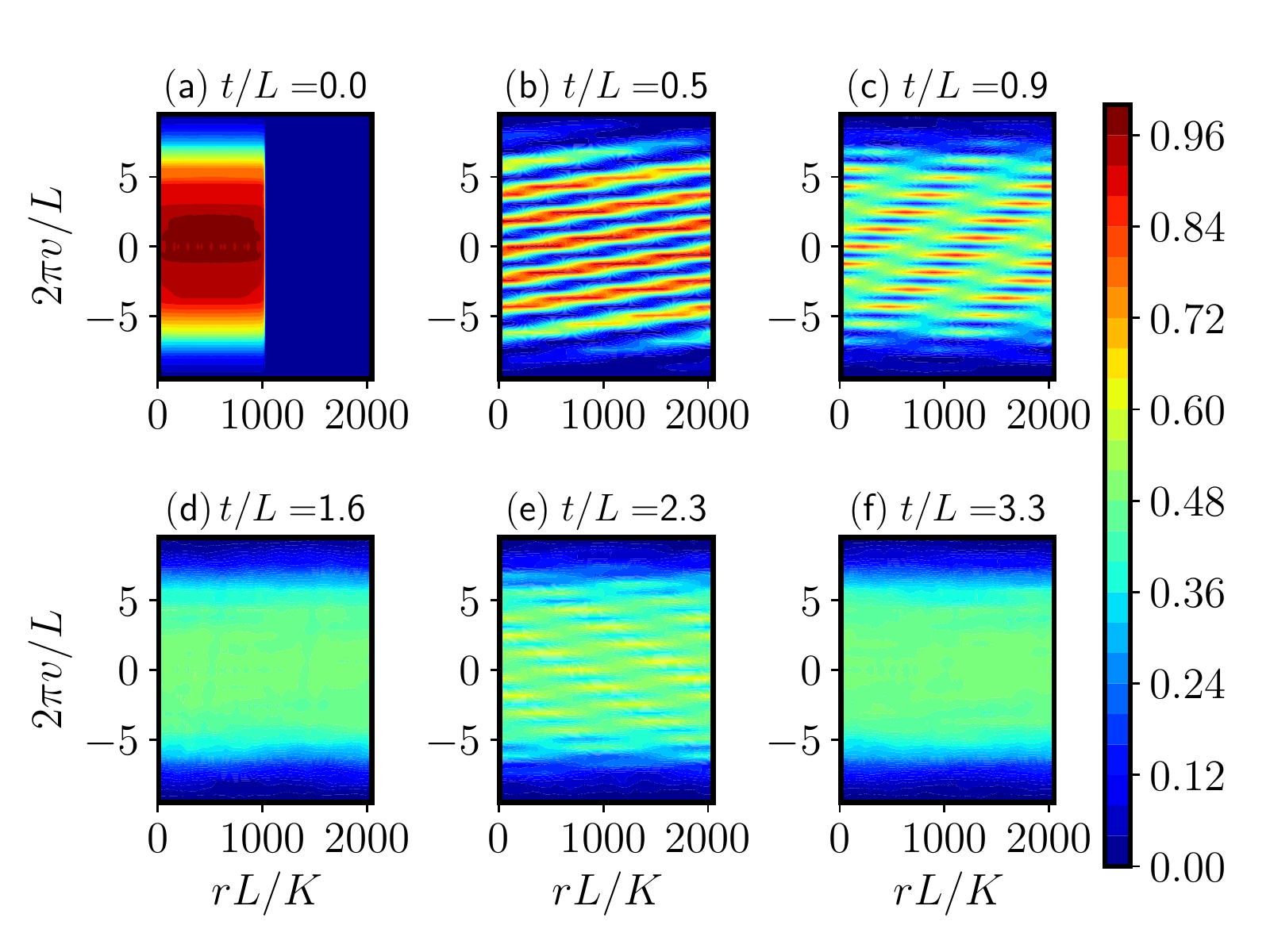}
}
\subfigure{
	\includegraphics[scale=0.45]{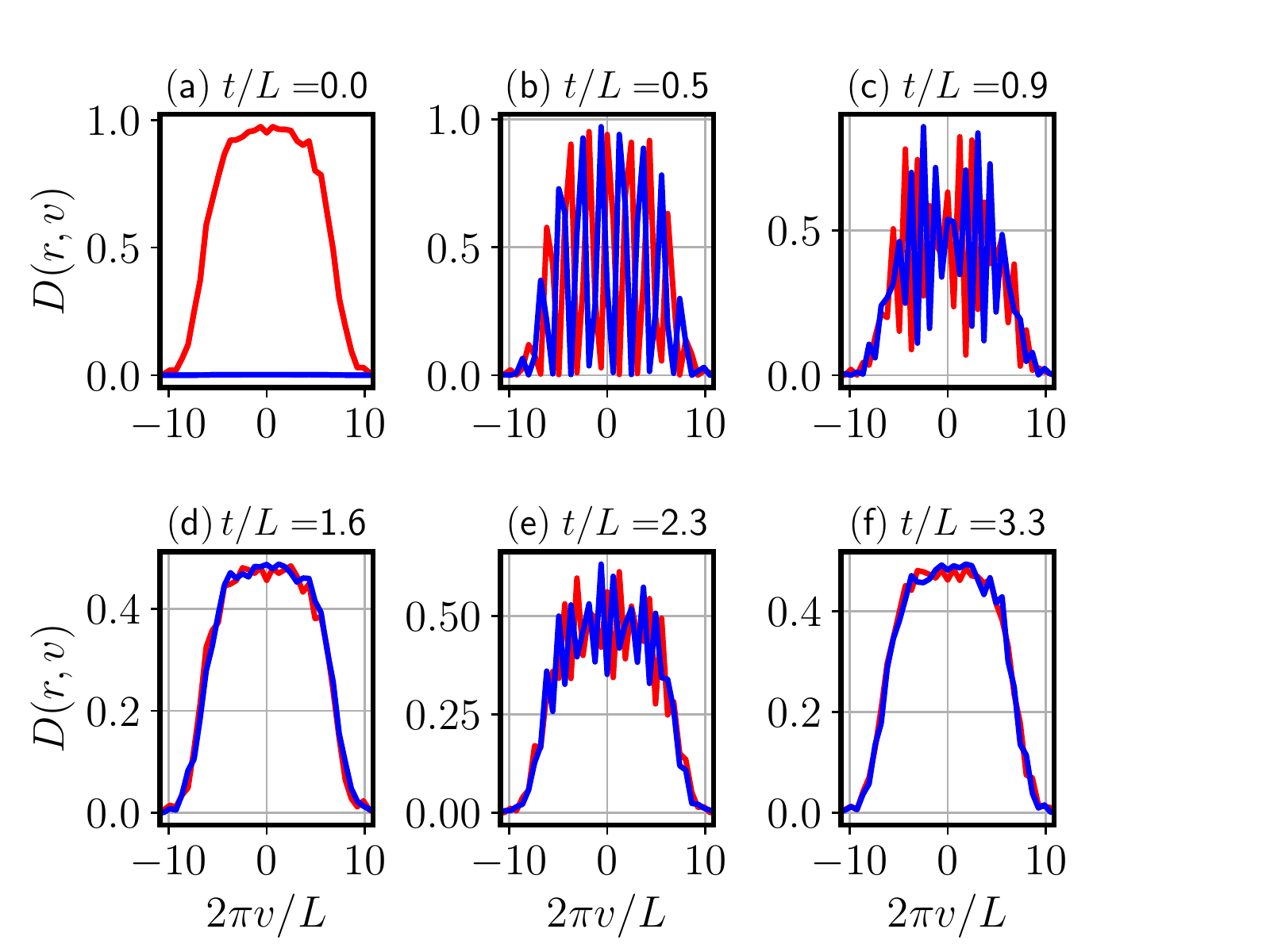}
}
\begin{minipage}{0.5\textwidth}
\subfigure{
	\includegraphics[scale=0.4]{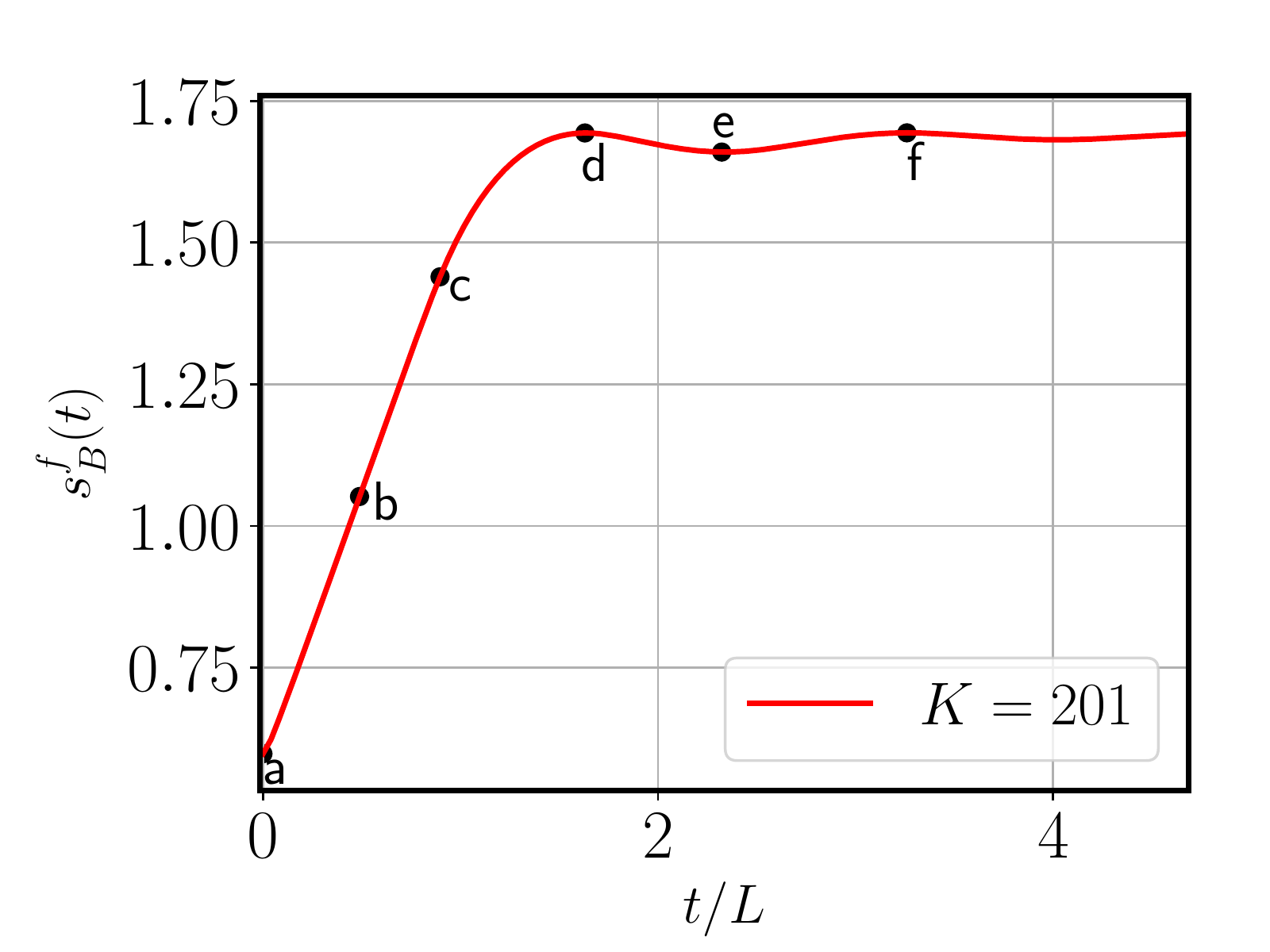}
}
\end{minipage}
\begin{minipage}{0.45\textwidth}
\caption{{\bf Fermions - evolution of $f$-macrovariables.} Top left: Heat map plot of the wave packet density  for parameters  $T=2\pi$,~$\mu=21.53$,~$N=L=2048$, and $K=201$. Top right:  The wave packet density as a function of momentum at two values of  $r=K/4$ (red) and $r=3K/4$ (blue), at the same time instances as in the heat map.  
Bottom left: The entropy evolution with time where the points $a-f$ correspond to the same time snapshots as in the top row. The relative values of the entropy at these points  are consistent with the presence of structures (or lack thereof) in the heat map and the cross-sectional profiles.}
\label{fFWPDfig}
\end{minipage}
\end{figure*}

\begin{figure*}

\subfigure[$T=2\pi/5$,~$\mu=19.82$,~$\Delta s = 1.28$]{
	\includegraphics[scale=0.43]{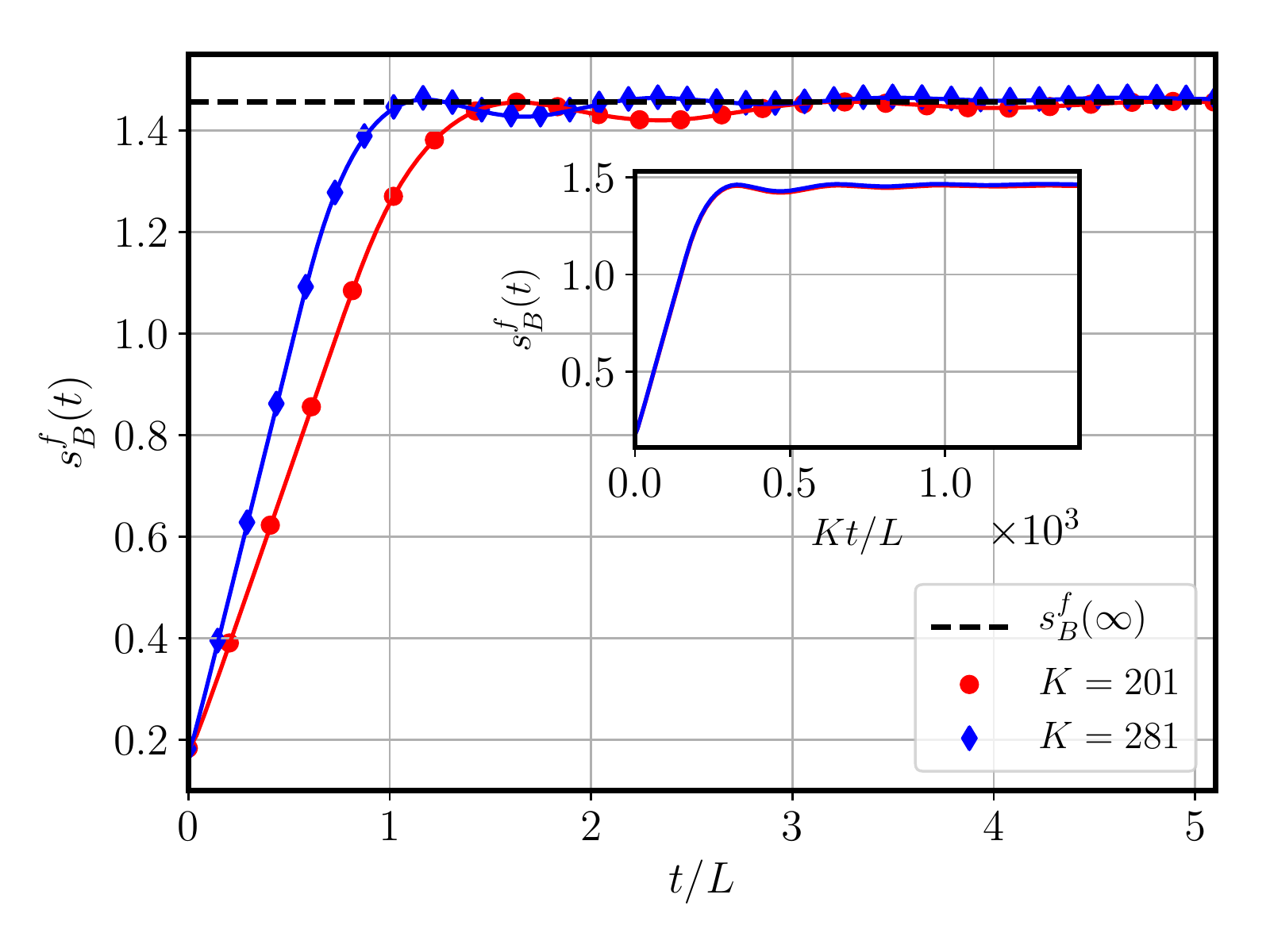}
}
\subfigure[$T=10\pi$,~$\mu=22.01$,~$\Delta s = 0.84$]{
	\includegraphics[scale=0.43]{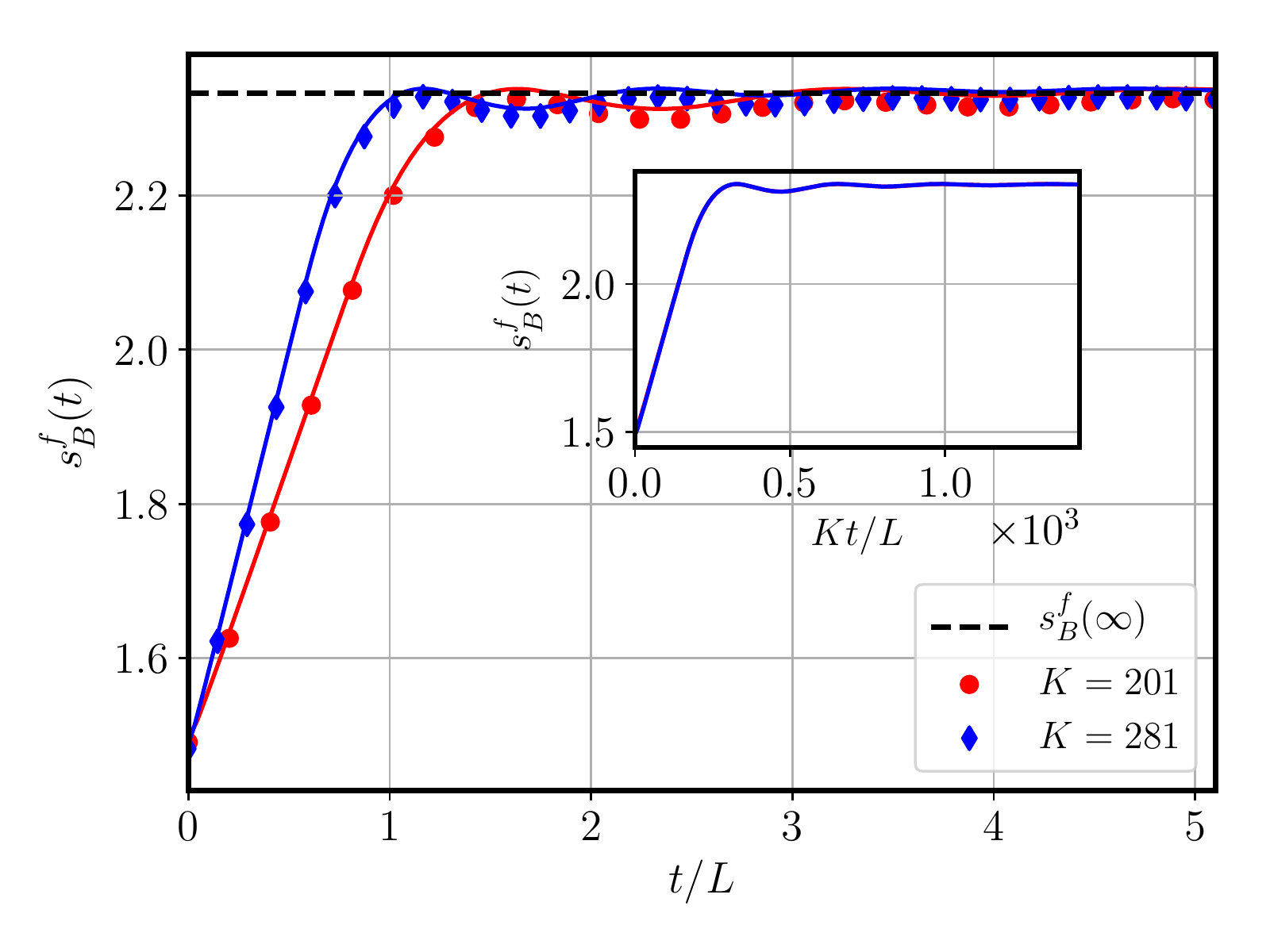}
}
\caption{{\bf Fermions}: Time evolution of the $f$-macrostate entropy per particle, $s_B^f$, at two different temperatures and for two coarse-graining scales $K$ at each temperature. We see very good agreement between results for the pure state (red dots) and  thermal state (solid lines) initial conditions. The inset in each plot shows the collapse of different $K$ curves on re-scaling time.} \label{SBffplots}
\end{figure*}

\section{Numerical results}
\label{sec:numerics}
In this section, we present the results of the evolution of the two macrovariables $f$ and $U$, and the corresponding entropies for fermions in Sec.~\eqref{subsec:fermions} and for bosons in Sec.~\eqref{subsec:bosons}. 

The equilibrium state of an ideal gas is described by its temperature $T$ and density $\rho=N/L$. For a quantum system, a relevant parameter that tells us whether we are in the quantum or classical regime is the ratio  of the interparticle distance $1/\rho$ to the  thermal de Broglie  wavelength, $\lambda_{\rm th}=h/(2 \pi m k_B T)^{1/2}$. With our choice of units with $m = k_B=\hbar=1$ and density fixed at $\rho=1$, we take the square of this ratio, $(\lambda_{\rm th} \rho)^{-2} =T/(2 \pi)$ to be the controlling parameter. In the following, we will present results for two sets of parameters: (i) low-temperature highly quantum regime $T=2 \pi/5$; (ii) high-temperature regime $T=10 \pi$. { In Fig.~\eqref{fermi-bose} we show the mean energy-level occupation number and their fluctuations for fermions and bosons at different temperatures.}  

We recall the two different initial conditions [see Sec.~\eqref{subsec:den_mat}] for which we present our results. 
\begin{enumerate}
\item  We consider the initial state to be a single pure many-body Fock state given by the box eigenstate: 
\begin{align}
|\Phi \ra =|\{ n_s \} \ra~, \end{align}
where the single-particle level box occupancies, $\{n_s\}$, are chosen with probabilities given by the grand-canonical distribution [in the box $x\in (0, L/2)$] with $T$, $\mu$ fixed at the desired values corresponding to our specified initial $T$, $\rho$.  We sample only one initial state this way; for large $N$, this comes close to ``self-averaging".  
\item We consider an initial state whose one-particle density matrix is identical to that of
the grand-canonical distribution in the box. 
\end{enumerate}

\subsection{Fermions}
\label{subsec:fermions}
In this section, we present results for fermions first for the $U$-macrovariable and then for the $f$-macrovariable.

\subsubsection{$U$-macrostate}
\noindent
In this case, we set $L=N=1024$ in all our numerics. Some of our main observations are:
\begin{enumerate}[(i)]
\item In Fig.~\eqref{2NPEfvsx} we show the spatial profiles of the $U$-macrovariables, given by the expectation values of the operators in Eq.~\eqref{U-var}, at different times starting with the gas in the region $(0, L/2)$ and with $A=20$  partitions of the system.   
Results are shown for both the  pure state and the thermal state initial conditions and we find a very good agreement between the two. We observe that at late times, all three fields approach uniform profiles which characterize our equilibrium state. 

\item   In Fig.~\eqref{fNvst}
 we consider two fixed cells centered at $x=L/4$ and $3L/4$ and plot the time-evolution of the number of particles in these cells. We again see a good agreement between the pure state and the thermal state initial conditions. An oscillatory relaxation to the uniform equilibrium state is observed. We note that the oscillation period in the low-temperature limit is given by $\tau_p = L/v_{\rm f}$ where $v_{\rm f}$ is the Fermi velocity. The amplitude of the oscillations decreases with increasing temperature. 
 \item The evolution of the Boltzmann entropy $S^U_B$ 
at the two temperatures are shown in Fig.~\eqref{SBUfplots},  for two choices of cell sizes, with $A=20,40$. We see
 a convergence of the growth curve with decreasing cell size.
A monotonic growth of the entropy and an eventual saturation to the expected equilibrium value (corresponding to uniform values of the conserved fields) is observed.
In the low-temperature case,  we observe an initial jump in the $S^U_B$ followed by a small flat regime and then a sharp increase. The initial jump size is smaller for finer coarse-graining.
On the other hand, the initial flat profile seen in Fig.~\eqref{SBUfplots}(a) is a result of the sharp cutoff in the momentum distribution in the low-temperature Fermi gas. The flat region is observed till time $(L/4)/v_{\rm f}$, where $v_{\rm f}$ is the Fermi velocity, which corresponds to the time taken by the gas to fill the circle for the first time.
\end{enumerate}

\subsubsection{$f$-macrostate}
\label{subsec:fm}

In this case, we set $L=N=2048$ in all our numerics. Some of our main observations are:
\begin{enumerate}[(i)]

\item To see the evolution of the $f$-macrovariables, we plot heat maps in the top left of Fig.~(\ref{fFWPDfig}) showing the values of wavepacket density $D_\alpha$ in the two-dimensional $\alpha=(r,v)$ plane  for an intermediate temperature $T=2 \pi$. We present six different time snapshots indicated by (a)-(e). 
The wavepacket density is initially flat in the box and there is an oscillatory relaxation to an eventual  flat profile over the circle. We see an interference pattern that disappears in (d) and reappears in (e) and then again disappears in (f). This feature is more manifest in  the top right  panel of Fig.~(\ref{fFWPDfig})  which shows  the wave packet density as a function of momentum at two values of  $r=K/4$ (red) and $r=3K/4$ (blue), at the same time instances as in the heat map. 
     These features lead to an  oscillatory relaxation of the entropy growth curve as seen in the bottom left panel of Fig.~(\ref{fFWPDfig}) where we find that the time for the entropy to first reach the saturation value, i.e,  the point (d) is given by $L^2/(2 \pi K)$. This is also the period of the subsequent oscillations seen in the entropy evolution.  This time scale can be understood within a semi-classical framework and using the results from Ref.~\cite{chakraborti2021entropy}. There it was shown that the time scale of oscillations of $s^f_B$ is given by $\tau=L/\Delta  v$, where $\Delta v$ is the momentum coarse-graining scale. In the quantum case, we replace this by $\Delta v= \Delta p/m= 2 \pi \hbar K/(mL)$. With our choice of units $\hbar=m=1$, we therefore get  $\tau=L^2/(2 \pi K)$. 
    
     \item In Fig.~(\ref{SBffplots}), we plot the entropy growth for the two temperatures (low and high) and for two different $K$ values corresponding to each temperature. At both coarse graining scales, $K$, we see a good agreement between the entropy calculated from the pure state and the thermal state initial conditions. The entropy saturates to a value as predicted from Eqs.~(\ref{wp entropy}, \ref{wp density inf 2}). In the inset,  we see a collapse of the data for different $K$ on scaling  time by $K$, consistent with the expression for $\tau$ mentioned above.  Note that similar oscillations and collapse of data were seen in the entropy growth in the free expansion of the non-interacting classical ideal gas~\cite{chakraborti2021entropy}.

\end{enumerate}

\subsection{Bosons}
\label{subsec:bosons}

We now present results for bosons for the $U$ and $f$ macrovariables. Except at very low temperatures, we find that several features are the same as that of fermions irrespective of the difference in statistics. We also highlight the striking differences between bosons and fermions. For bosons at very low temperatures, only levels with the energy of order $\lesssim k_BT$ will be occupied. The number of occupied levels is thus small unless we consider very large $N$. Therefore it is difficult to see relaxation at such low temperatures and we do not consider such temperatures. 

\begin{figure*}[h]
\centering
\subfigure[$T=2\pi/5$,~$\mu=-0.11$]{
	\includegraphics[scale=0.3]{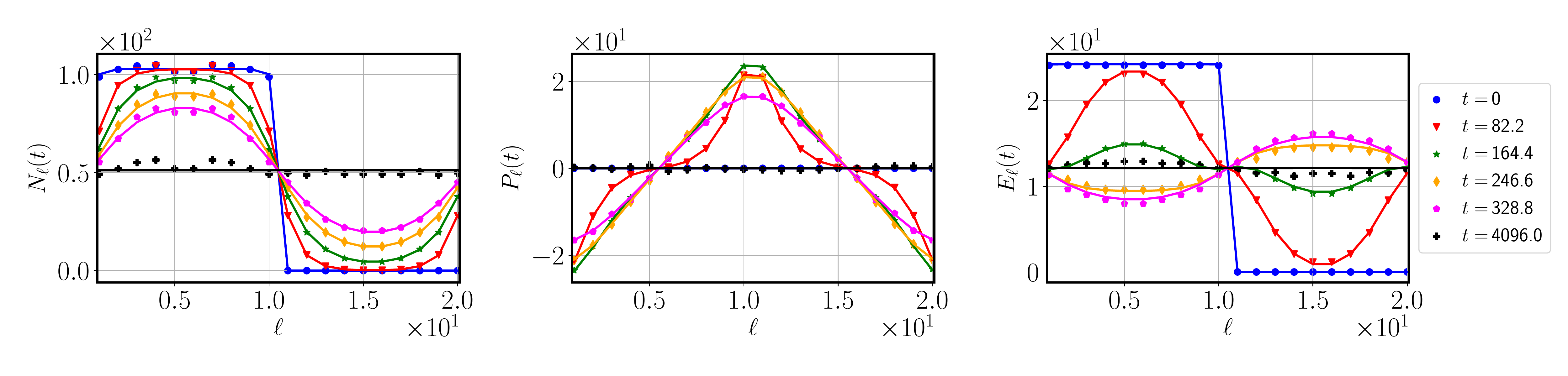}
}
\subfigure[$T=10\pi$,~$\mu=-19.86$]{
	\includegraphics[scale=0.3]{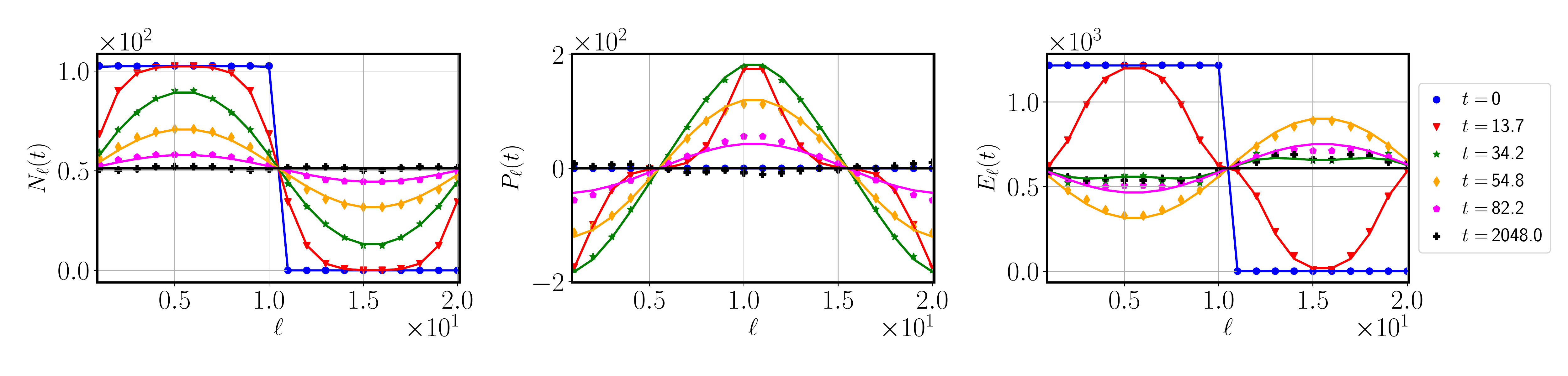}
}
\caption{{\bf Bosons - evolution of $U$-macrovariables}: Plots showing the spatial profiles  of the number of particles, the total momentum, and the total energy in each of the $A = 20$ cells for $N=L=1024$ at different times. Results are presented for  two different  temperatures $T=2 \pi/5,~10 \pi$, and chemical potentials are fixed so that the mean density is set at $\rho=1$. We see a reasonable agreement between the pure state (red dots) and the thermal state (solid lines) results, but significant deviations (due to finite-size effects) are observed at the latest times. 
 \label{2NPEbvsx}}
\end{figure*}

\subsubsection{$U$-macrostate}
In Fig.~(\ref{2NPEbvsx}) we show the spatial profiles of the $U$-macrovariables, given by the expectation values of the operators in Eq.~(\ref{U-var}), at different times starting with the gas in the left half of the box. We again consider a partition of the box into $A=20$ cells. Results are shown for both the pure state and the thermal state initial conditions and as before we find reasonable agreement between the two though we find significant differences at long times. In general, we find that for bosons the agreement is not as good as that of fermions due to stronger finite size effects. This is because, for a fixed total particle number and temperature, the typical number of occupied levels in the pure state is less for bosons compared to that for fermions and fluctuations are larger~[see Fig.~\eqref{fermi-bose}].

\begin{figure*}
\subfigure[$T=2\pi/5$,~$\mu=-0.11$,~$N=L=1024$]{
	\includegraphics[scale=0.43]{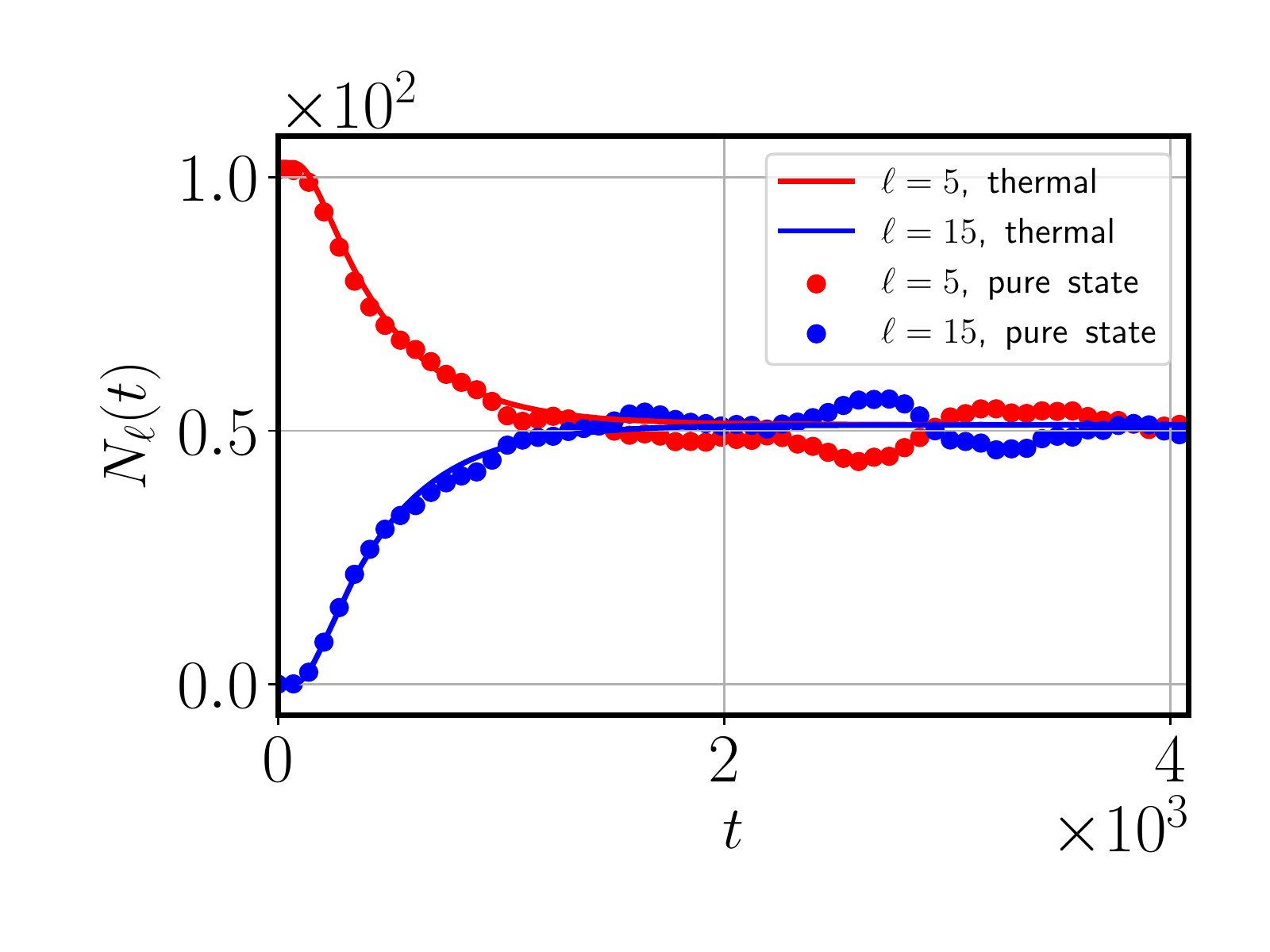}
}
\subfigure[$T=10\pi$,~$\mu=-19.85$,~$N=L=1024$]{
	\includegraphics[scale=0.43]{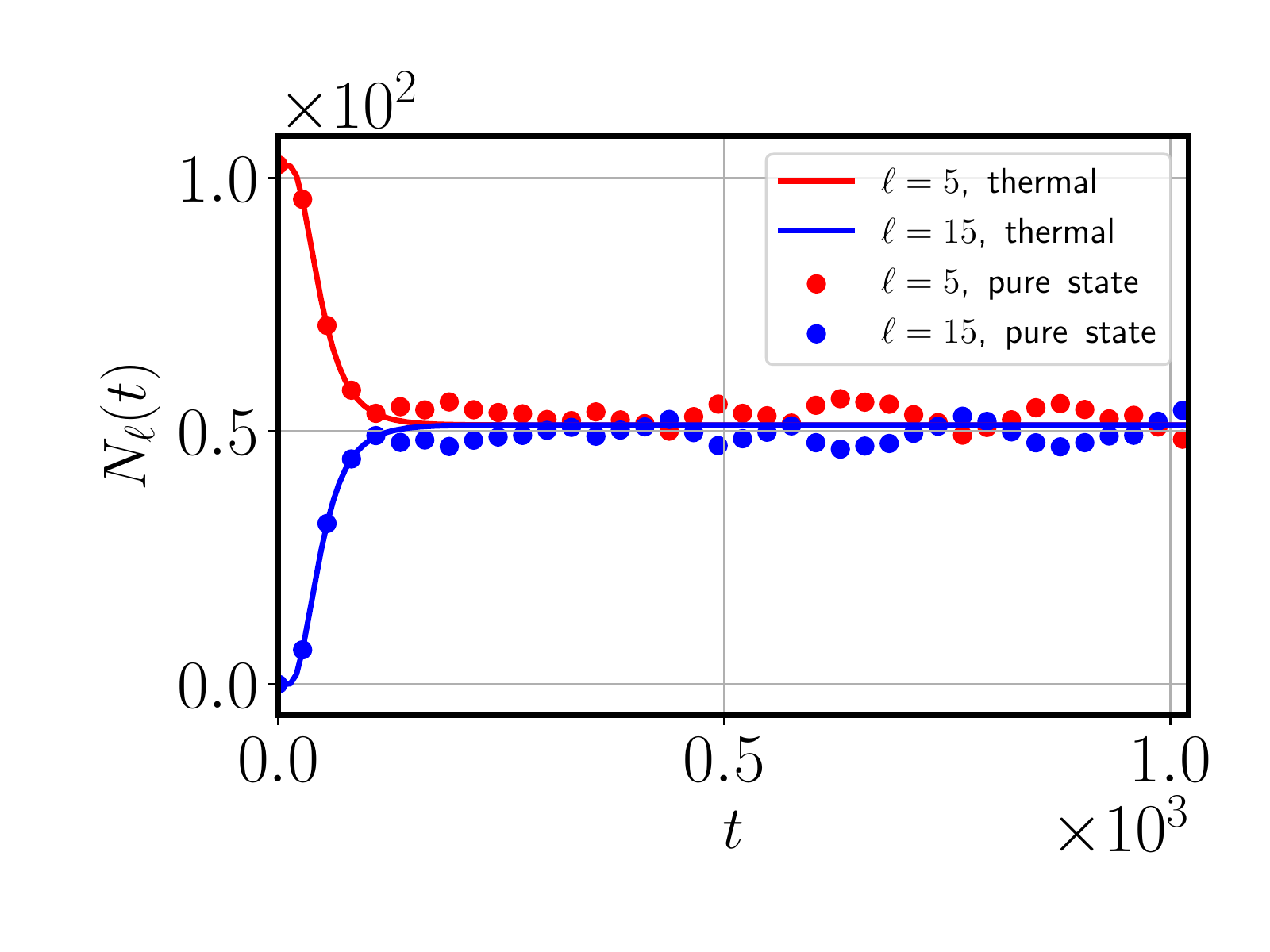}
}
\caption{{\bf Bosons}: Time evolution of the number of particles inside the $5^{th}$ and $15^{th}$ cells out of a total of $A = 20$ cells for $N = L = 1024$. As in Fig.~\eqref{2NPEbvsx},  we see a reasonable agreement, at early times, between the pure state (dots) and the thermal state results (solid line) while at longer times there are significant deviations.  Note that the thermal data shows the absence of  oscillations while in the pure state data, we see fluctuations that are expected to decrease with increasing system size.
 \label{bNvst}}
\end{figure*}

\begin{figure*}[t]

\subfigure[$T=2\pi/5$,~$\mu=-0.11$,~$\Delta s = 0.31$]{
	\includegraphics[scale=0.43]{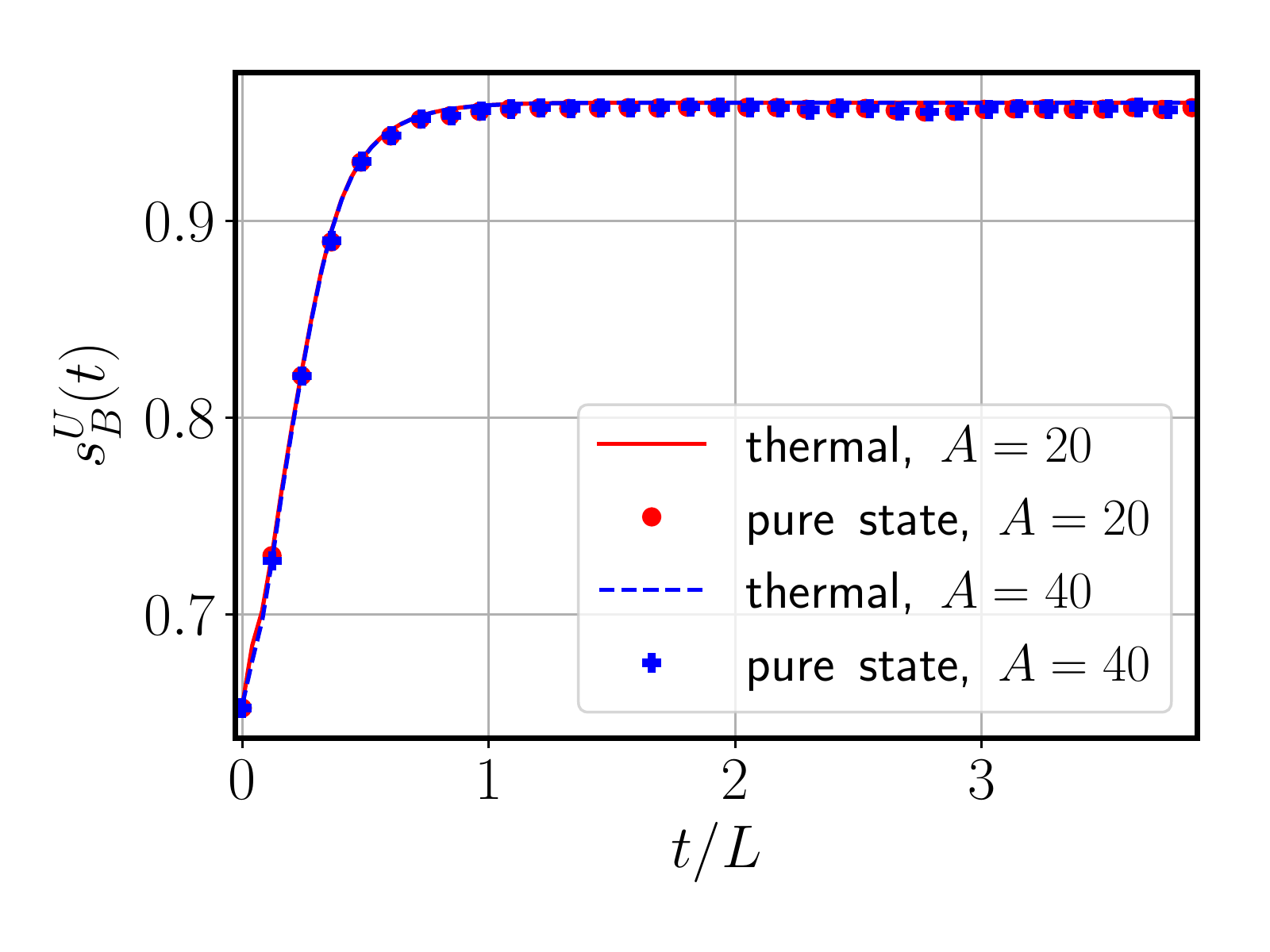}
}
\subfigure[$T=10\pi$,~$\mu=-19.86$,~$\Delta s = 0.56$]{
	\includegraphics[scale=0.43]{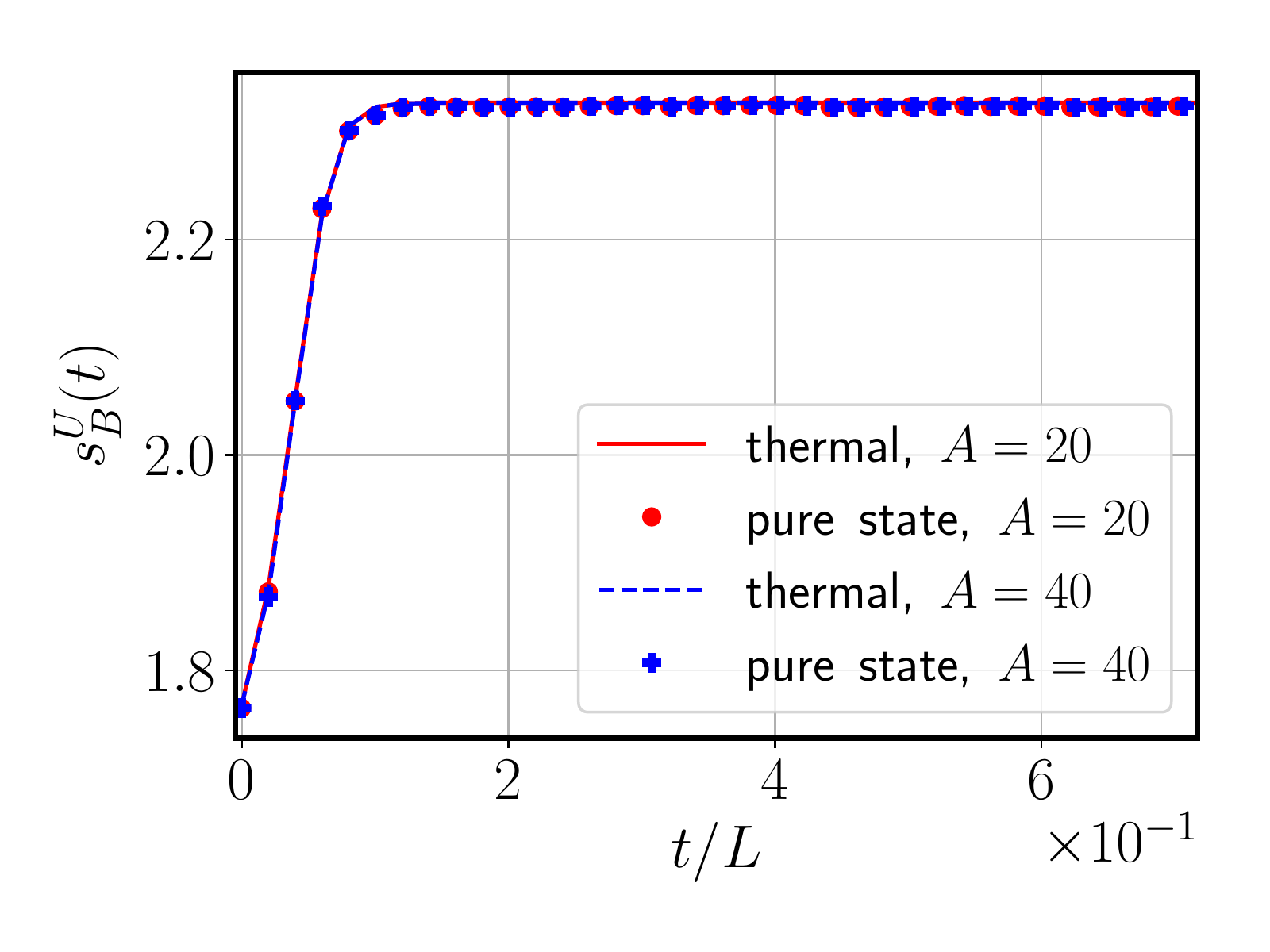}
}
\caption{{\bf Bosons}: 
$U$-macrostate entropy growth  for $N = L = 1024$ at two different temperatures and for two different coarse-graining scales with $A=20$ (blue lines) and $A=40 $ (red lines). Results are presented for the pure state (dots) and  thermal state (solid lines) initial conditions.  Unlike for fermions, the initial sharp rise is not seen in the case of bosons. At large times, in all cases, the entropy saturates to the thermodynamic entropy of the new equilibrium state (corresponding to uniform values of the three conserved fields on the circle). The sub-captions give the values for the change in entropy per particle, $\Delta s$, for the two cases.} 
\label{SBUbplots}
\end{figure*}

In Fig.~\eqref{bNvst}
 we consider two fixed cells located on the two halves of the circle centered at $x=L/4$ and $3L/4$ and plot the time-evolution of the number of particles inside these cells. We again see a reasonable agreement between the results from the pure and thermal states, though the differences are significantly larger than what was seen for fermions. 
{It is worth noting that unlike in the case of fermions, here we do not see any oscillations but rather a monotonic relaxation to the uniform equilibrium state.   The oscillations in the fermionic case arise due to the sharpness of the  distribution near the Fermi energy at low temperatures which allows one to define a typical velocity $v_F$ -- hence a period of oscillation $\tau_p=L/v_F$. On the other hand in the case of bosons, we cannot identify such a typical velocity. Note that even though the evolution equation for the Wigner function is formally identical for bosons and fermions (as also classical particles),   differences arise due to the form of the initial conditions. More precisely, for a typical pure state corresponding to thermal equilibrium, the form of the initial state is entirely different for bosons and fermions.}
 The evolution of the Boltzmann entropy per particle $s^U_B$ at the two temperatures is shown in Fig.~\eqref{SBUbplots},  for two choices of cell sizes  $A=20,40$. We observe: (i)  a convergence of the growth curve with decreasing cell size; (ii) a monotonic growth of the entropy and an eventual saturation to the expected equilibrium value (corresponding to uniform values of the conserved fields).

\begin{figure*}
\subfigure{
	\includegraphics[scale=0.5]{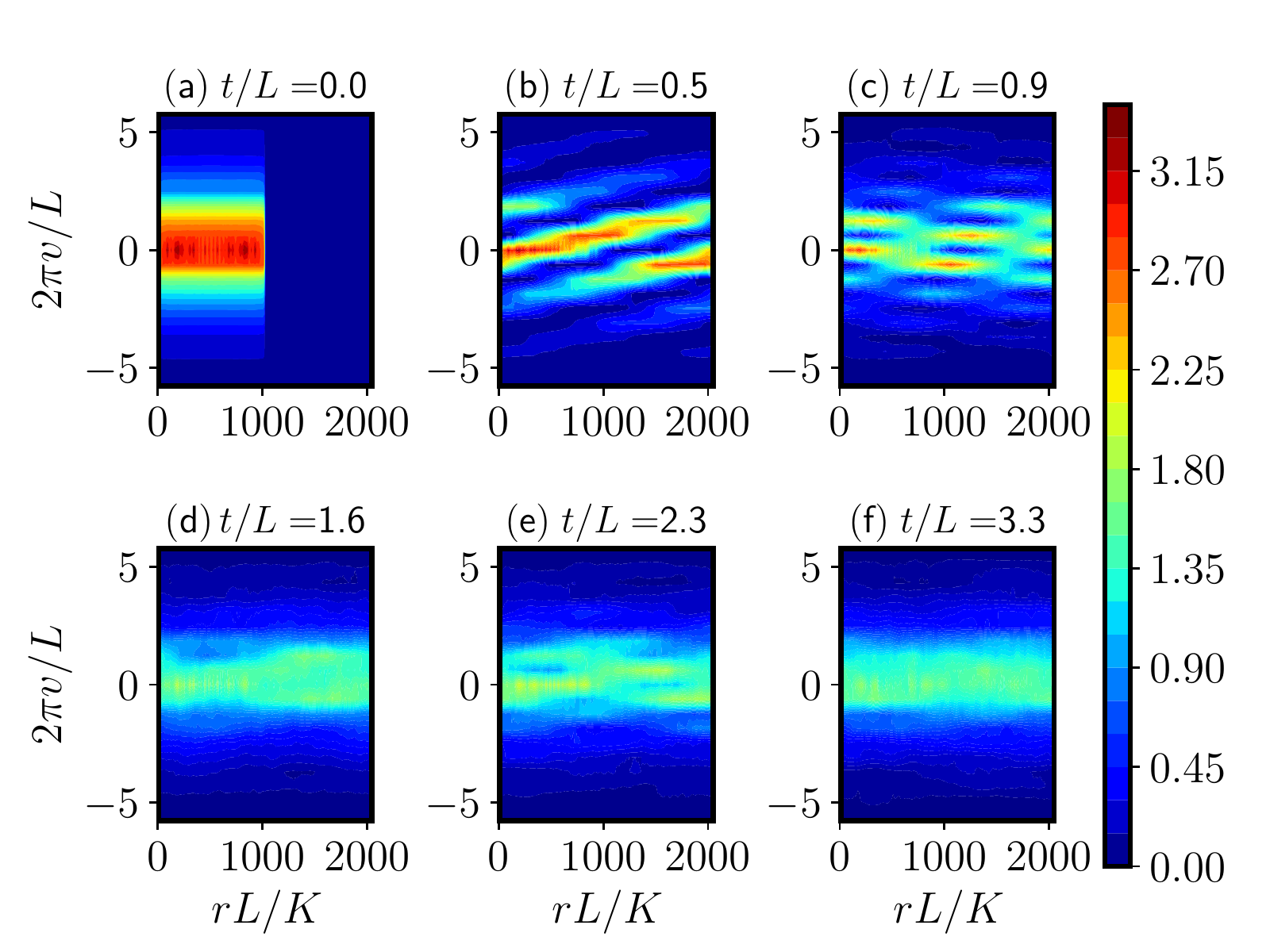}
}
\subfigure{
	\includegraphics[scale=0.5]{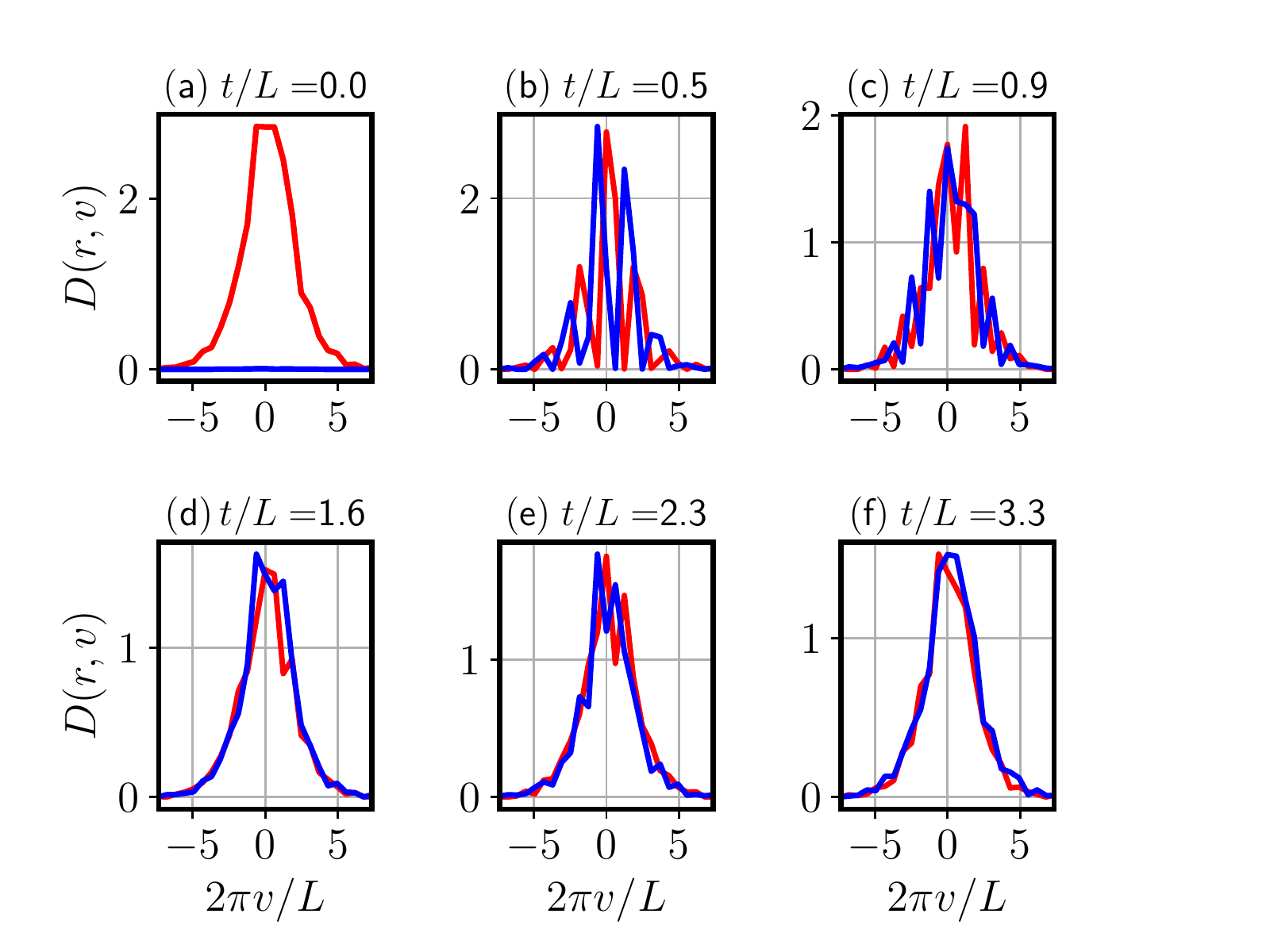}
}
\begin{minipage}{0.45\textwidth}
\subfigure{
	\includegraphics[scale=0.4]{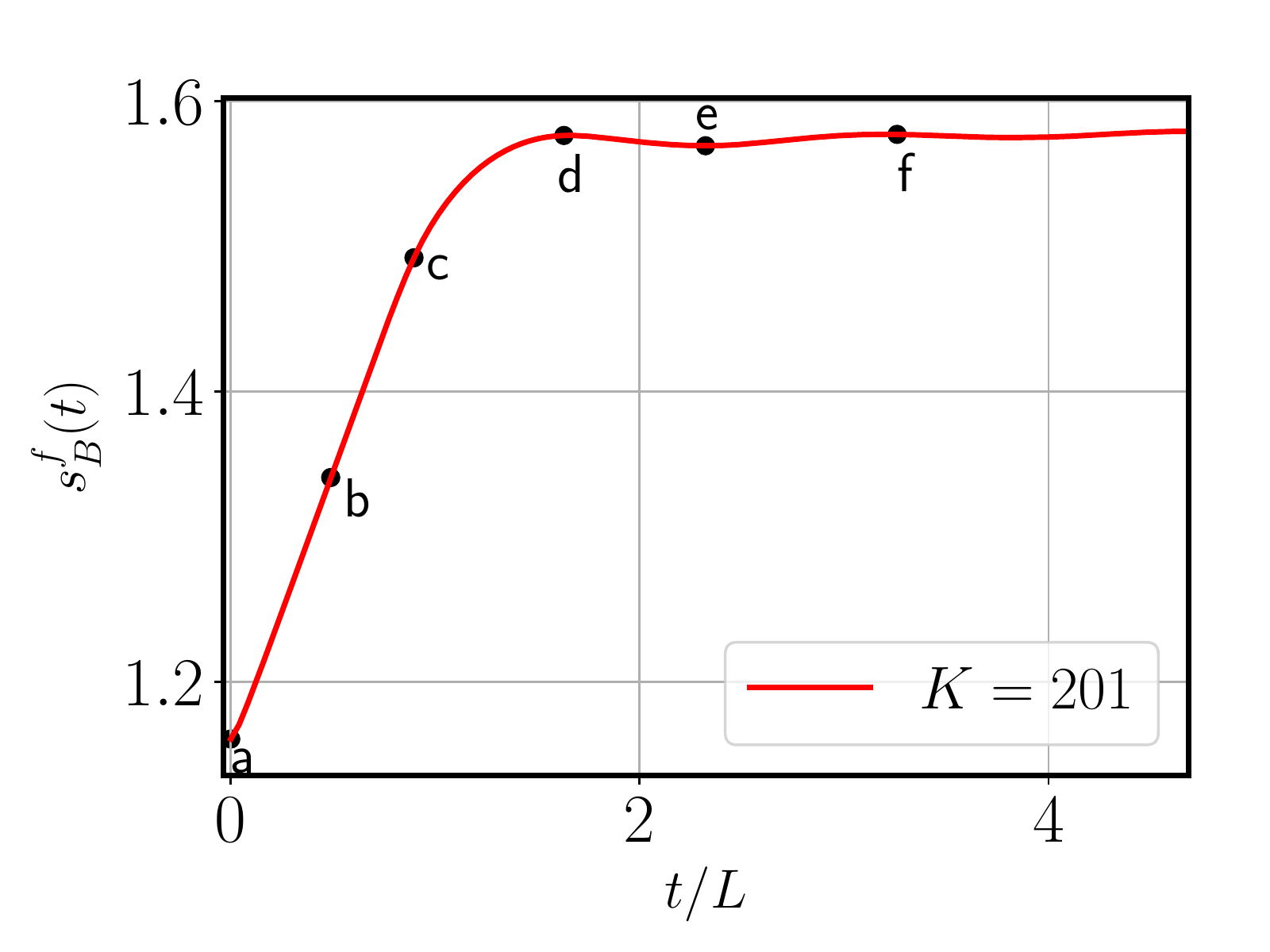}
}
\end{minipage}
\begin{minipage}{0.5\textwidth}
\caption{{\bf Bosons -   evolution of $f$-macrovariables.} Top left: Heat map plot of the wave packet density at $T=2\pi$,~$\mu=-1.7$,~$N=L=2048$, and $K=201$. Top right: The wave packet density as a function of momentum at two values of  $r=K/4$ (red) and $r=3K/4$ (blue), at the same time instances as in the heat map. Bottom left: The entropy evolution with time where the points $a-f$ correspond to the same time snapshots as in the top row. Like fermions, here also the relative values of the entropy at these points are consistent with the presence of structures (or lack thereof) in the heat map and the cross-sectional profiles.}
\label{bFWPD}
\end{minipage}
\end{figure*}

\begin{figure*}
\subfigure[$T=2\pi/5$,~$\mu=-0.11$,~$\Delta s = 0.29$]{
	\includegraphics[scale=0.43]{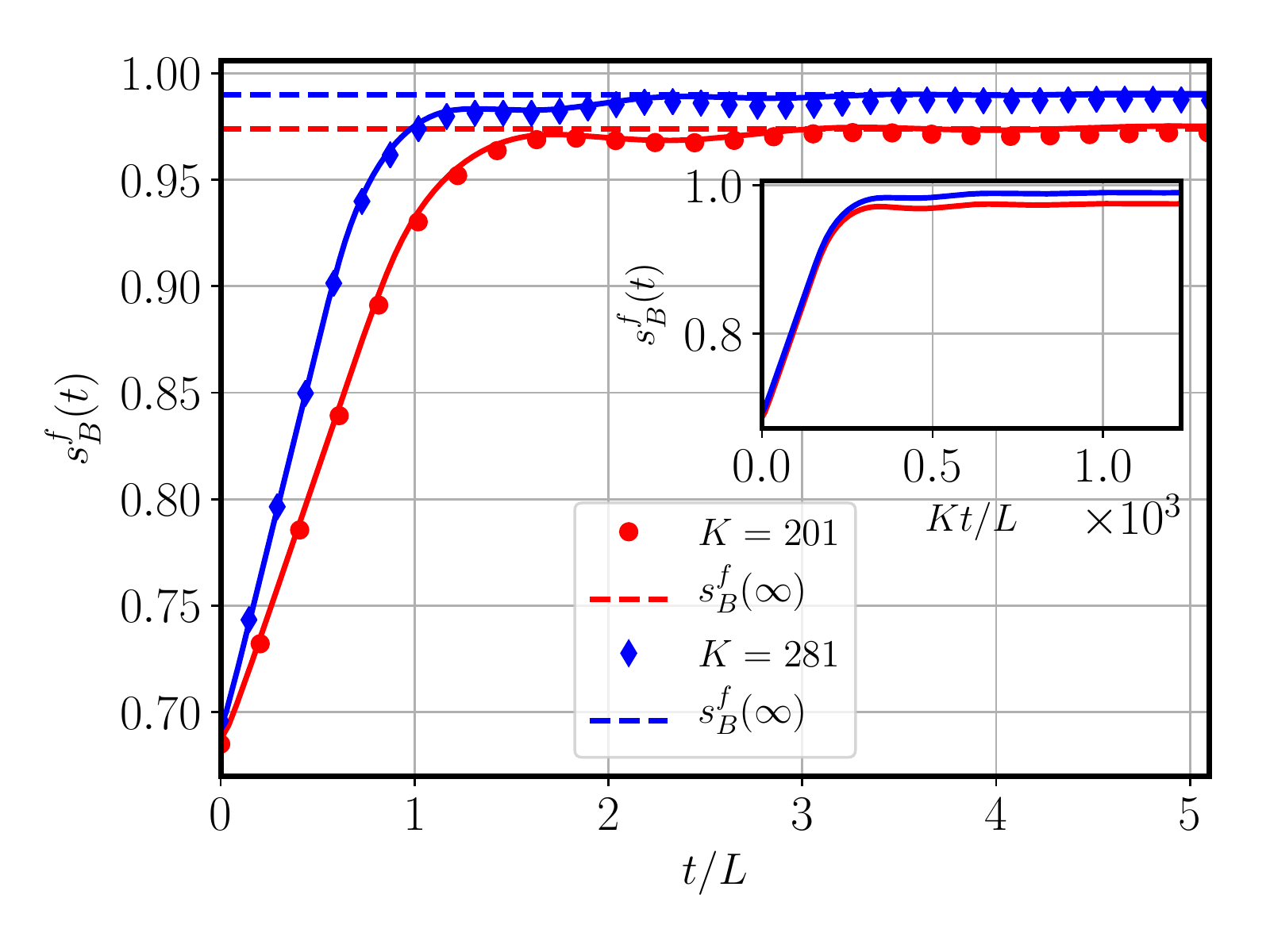}
}
\subfigure[$T=10\pi$,~$\mu=-19.86$,~$\Delta s = 0.53$]{
	\includegraphics[scale=0.43]{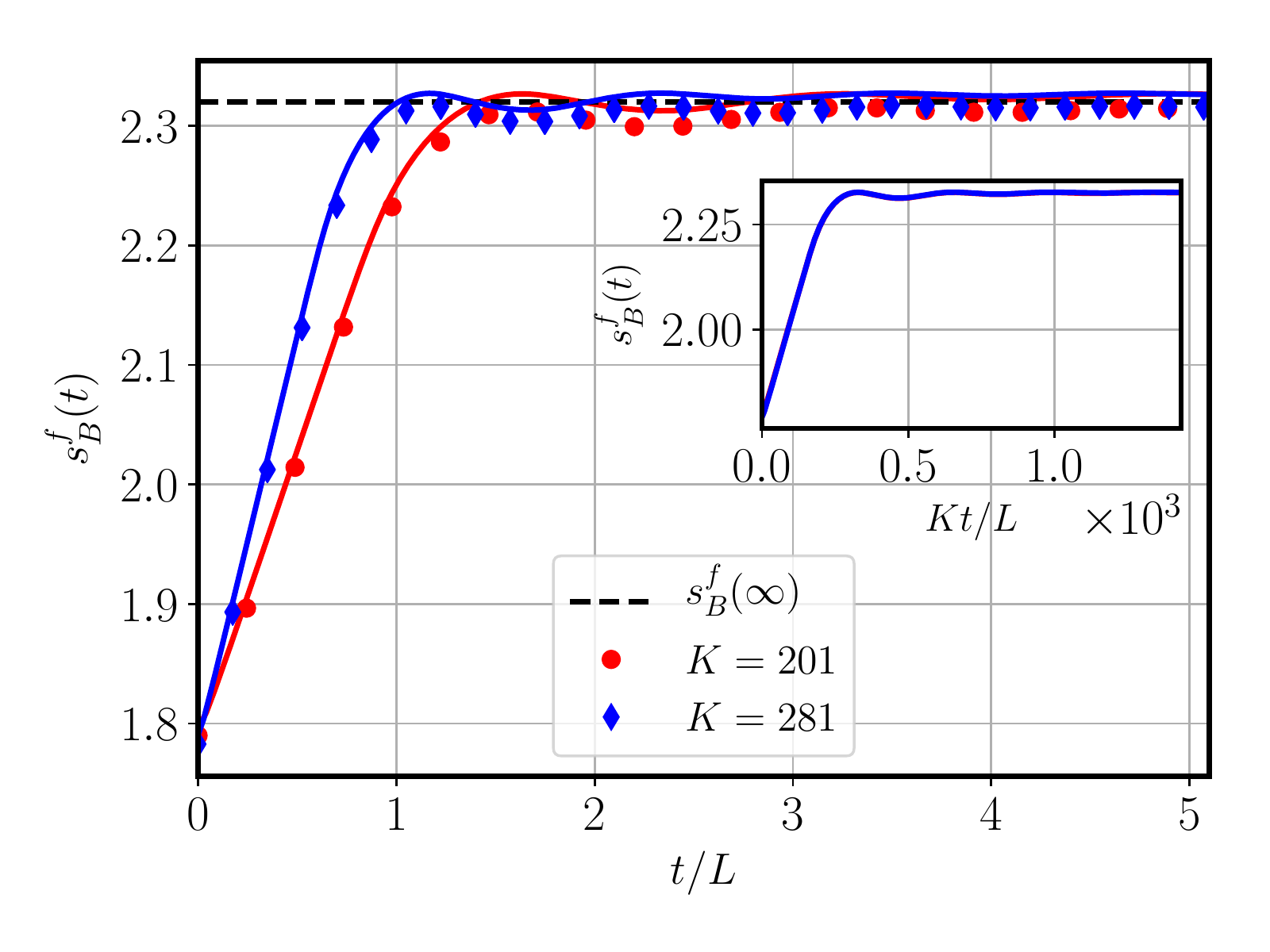}
}
\caption{{\bf Bosons}: Time evolution of the $f$-macrostate entropy per particle, $s_B^f$, at two different temperatures and for two coarse-graining scales $K$ at each temperature. We see a good agreement between results for the pure state (red dots) and  thermal state (solid lines) initial conditions. The inset in each plot shows the collapse of different $K$ curves on re-scaling time.} \label{SBUfplots2048}
\end{figure*}

\subsubsection{$f$-macrostate}
For the $f$-macrostate, we again plot the heat map of the wavepacket density and the corresponding entropy growth curve in Fig.~\eqref{bFWPD}. The results look similar to fermions, except for the fact that the wavepacket density for bosons is more smeared out as compared to fermions. This is due to the usual difference between the Bose and the Fermi function at low temperatures. However, in the entropy growth curves for bosons in Fig~\eqref{SBUfplots2048}, the saturation value of the entropy depends rather strongly on $K$, especially in the low-temperature case [see details in appendix \eqref{app:ONR}].

\section{Discussions and Conclusions}\label{sec:conclusion}
The main aim of this paper has been to use Boltzmann's ideas to construct an entropy function that can be defined for a pure quantum state and which allows us to characterize irreversibility in macroscopic systems. 
For the example of the quantum ideal gas, two sets of macroscopic descriptions (called $U$ and $f$) were defined that provide a coarse-grained view of the system, which is in a pure quantum state.  The evolution of the entropy functions associated with these macrostates was studied for the case where the gas, initially in a pure state and spatially confined, was allowed to expand to twice its volume. 
We summarize and comment on some of our main results.

\begin{enumerate}
    \item  {The $U$-macrovariables are the coarse-grained operators corresponding to the number, momentum, and energy of particles in spatial cells of size $\delta = L/A$. For both fermions and bosons, we see that the macrostates reach a steady state characterized by the three fields reaching homogeneous spatial profiles. Depending on the parameter values, we see either damped or oscillatory relaxation of the fields to the steady state. However,  we always observe a  monotonic increase of the associated entropy, $s^U_B$.}  
    \item  {The definition of our $f$-macrostate is aimed at obtaining an analogue of the single-particle phase space density in classical systems. A natural candidate for this are the number operators corresponding to a localized wavepacket set of basis states. While these {operators are not bonafide macrovariables, we nevertheless can use their {\it expectation values} as macrovariables to identify macrostates and compute an entropy function by maximizing the Gibbs-von Neumann entropy, given those expectation values. }}
    
 The wavepackets that we construct are located at discrete space and momentum points, denoted $\alpha\equiv (r,v)$, and localized  on a scale of $\hbar$. We show that the average occupancy of these states denoted $D_\alpha$, constitute a coarse-graining of the Wigner function on a scale of $\hbar$. Unlike the Wigner function, $D_\alpha$ is positive and is thus similar to the so-called Husimi function~\cite{husimi1940}. However, the Husimi function has the full information on the one-particle density matrix and so for non-interacting systems, it cannot be used for the construction of an entropy function. The function $D_\alpha$ incorporates coarse-graining required to demonstrate entropy growth 
 in the quantum ideal gas. We note that the Wehrl entropy~\cite{wehrl1979} uses the Husimi function and has been used to study irreversibility in interacting quantum systems~\cite{tsukiji2016,kunihiro2009,goes2020}. 
    
 Our $f$-macrostate entropy, $S^f_B$, also increases with time and reaches a steady state saturation value. 
 In this case, the entropy growth is oscillatory with a period given by $L/\Delta p$ where $\Delta p$ is the momentum coarse-graining scale --  this can be understood from semi-classical considerations. The change in entropy per particle satisfies the bound   $ \ln 2 \leq \Delta s^f_B \leq 2 \ln 2$ for fermions and  $0  \leq \Delta s^f_B \leq \ln 2$ for bosons. This can be understood from the momentum distribution in the final state.
    \item Results obtained for the evolution of pure states were compared with those of  corresponding thermal initial states and we showed evidence of their equivalence at larger system sizes. This demonstrates typicality in the dynamical evolution. Bosons showed larger finite size effects at low temperatures because of the fact that a relatively smaller number of single-particle levels are occupied and number fluctuations are larger. 
    \item   From the single-particle  spectrum on the circle, it is clear that the system has an exact recurrence at a time $\tau_{rec}=L^2/\pi$ (in units of $m = \hbar = 1$). However for both our macrovariables, {relaxation to the steady state} 
    occurs on a time scale $\tau_{\rm eq} \sim L$ and so, in the thermodynamic limit $L, N \to \infty$ with $L/N$ constant, there is a clear separation between the relaxation and recurrence time-scales. This kind of  recurrence  is expected whenever the single-particle spectrum is given in terms of integers and would be observed in quantum particles in harmonic traps but would not be present in generic anharmonic potentials (see for e.g ~\cite{dean2019}).
\end{enumerate}

  As expected, our system does not reach a Gibbs equilibrium state. The final effective temperature and chemical potential corresponding to the final particle and energy density do not determine the true single-particle momentum distribution, which is not able to attain thermal equilibrium for the quantum ideal gas.  This is different from the classical gas, where an initial Maxwell velocity distribution would continue to be the correct equilibrium velocity distribution for the expanded gas. In the quantum gas, it would be necessary to include interactions to allow the momentum distribution to relax to thermal equilibrium; this does not happen for the ideal quantum gas. A recent study considered Boltzmann's entropy growth in a classical interacting gas and interesting differences with the non-interacting case were noted~\cite{chakraborti2023}. The effect of interactions in the quantum case would be interesting to explore, however, this would then become a highly-entangled many-body system and thus very challenging to treat accurately.

\section{Acknowledgements}

M.K. would like to acknowledge support from the project 6004-1 of the Indo-French Centre for the Promotion of Advanced Research (IFCPAR), Ramanujan Fellowship (SB/S2/RJN-114/2016), SERB Early Career Research Award (ECR/2018/002085) and SERB Matrics Grant (MTR/2019/001101) from the Science and Engineering Re- search Board (SERB), Department of Science and Technology (DST), Government of India. A.K. acknowledges the support of the core research grant CRG/2021/002455 and the MATRICS grant MTR/2021/000350 from the SERB, DST, Government of India. A.D., M.K., and A.K. acknowledges support of the Department of Atomic Energy, Government of India, under Project No. 19P1112R\&D. D.A.H. was supported in part by (USA) NSF QLCI grant OMA-2120757.

\appendix

\section{Marginals of the wavepacket density} \label{app:marginals}
Here we establish the fact mentioned in Sec.~\eqref{sec:f_macro} that the marginals of the wave-packet density, $D_\alpha$, after integrating either over momentum or space,  correspond respectively to the coarse-grained particle density and momentum density. 
Let us consider the momentum marginal first i.e. sum over $r$
\begin{align}
\begin{split}
D_v(v, t) &= \sum_r D(r, v, t) = \sum_{r, m, n} \langle r, v | m \rangle \tilde{\rho}_1(p_m, p_n, t) \langle n | r, v \rangle \\
&= \frac{1}{K} \sum_{r=1}^K \sum_{m, n \in \mathcal{R}_v} e^{2\pi i (m-n) r/K} \tilde{\rho}_1(p_m, p_n, t) \\
&= \sum_{n \in \mathcal{R}_v} \tilde{\rho}_1(p_n, p_n, 0).
\end{split} \label{p marginal A}
\end{align}
Let us now consider the r-marginal i.e. sum over $v$
\begin{align}
&D_r(r, t) = \sum_v D(r, v, t) = \int \frac{dx dx'}{L K}\sum_v \sum_{m, n \in \mathcal{R}_v} Q_{m n}(x, x', t), \label{r marginal A} \\
&Q_{m n}(x, x', t) = e^{2\pi i(-m x+n x')/L} e^{i (m - n) 2\pi r/K} \rho_1(x, x', t). \label{Q_def A} 
\end{align}
Note that $m, n \in \mathcal{R}_v$ which is given by 
\begin{equation}
\mathcal{R}_v = \left\{ v-\displaystyle{\frac{K-1}{2}}, \ldots, v+\frac{K-1}{2} \right\}. 
\label{eq:app_rv}
\end{equation}
We substitute $m=v+\tilde m$ and $n=v+\tilde n$ to get rid of the $v$ dependence in the sums over $m$ and $n$. We thus get 
\begin{align}
D_r(r, t) = \int \frac{dx dx'}{L K} \sum_v \sum_{\tilde m, \tilde n \in \mathcal{R}_0} e^{2\pi i v(x'-x)/L} Q_{\tilde m  \tilde n}(x, x', t). \label{r marginal A 2}
\end{align}
Now since $v$ takes values $\vartheta K$ where $\vartheta$ is an integer that runs from $-\infty$ to $+\infty$, the sum over $v$ can be done and gives
\begin{align}
\begin{split}
D_r(r, t) &= \int \frac{dx dx'}{ K^2}\sum_{\tilde m, \tilde n \in \mathcal{R}_0} \delta(x-x') Q_{\tilde m \tilde n}(x, x', t) \\
&= \int\limits_0^L \frac{dx}{ K^2} \abs{\sum_{\tilde n= -(K-1)/2}^{(K-1)/2} \exp[ 2 \pi i \tilde n \left( \frac{x}{L} - \frac{r}{K} \right)]}^2 \rho_1(x,x, t) \\
&= \int\limits_0^L dx \,h_K\left( \frac{x}{L} - \frac{r}{K} \right) n(x, t),
\end{split} \label{r marginal A 3}
\end{align}
where $n(x, t)$ is the particle density and $h_K(x) = \displaystyle{\frac{1}{K^2}} \left( \frac{\sin \pi K z}{\sin \pi z} \right)^2$.

\section{Dependence on $K$}
\label{app:ONR}

\begin{figure}[h]
\centering

\subfigure[$T = 2 \pi/5$]{
	\includegraphics[scale=0.4]{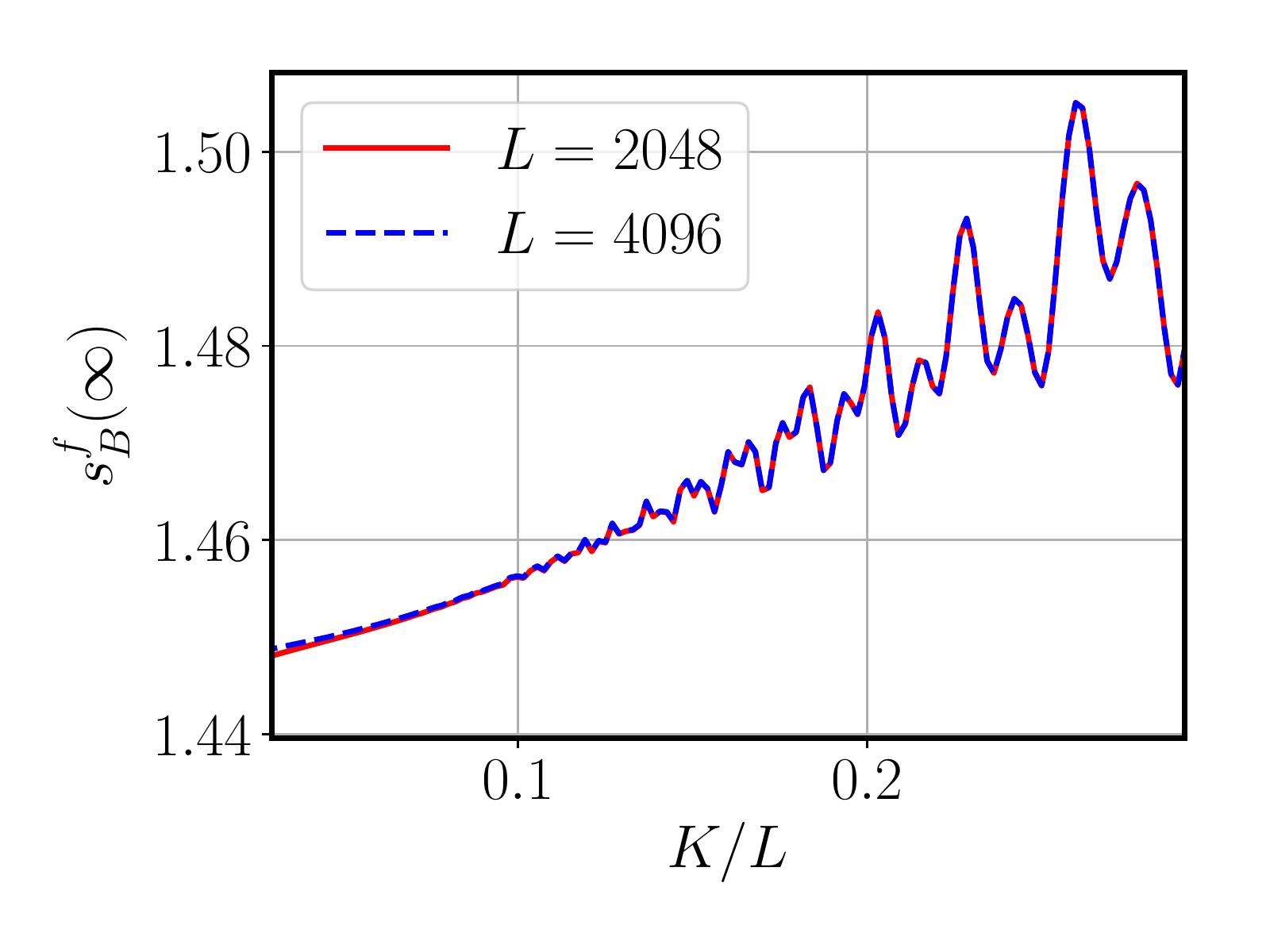}
}
\subfigure[$T = 2 \pi$]{
	\includegraphics[scale=0.4]{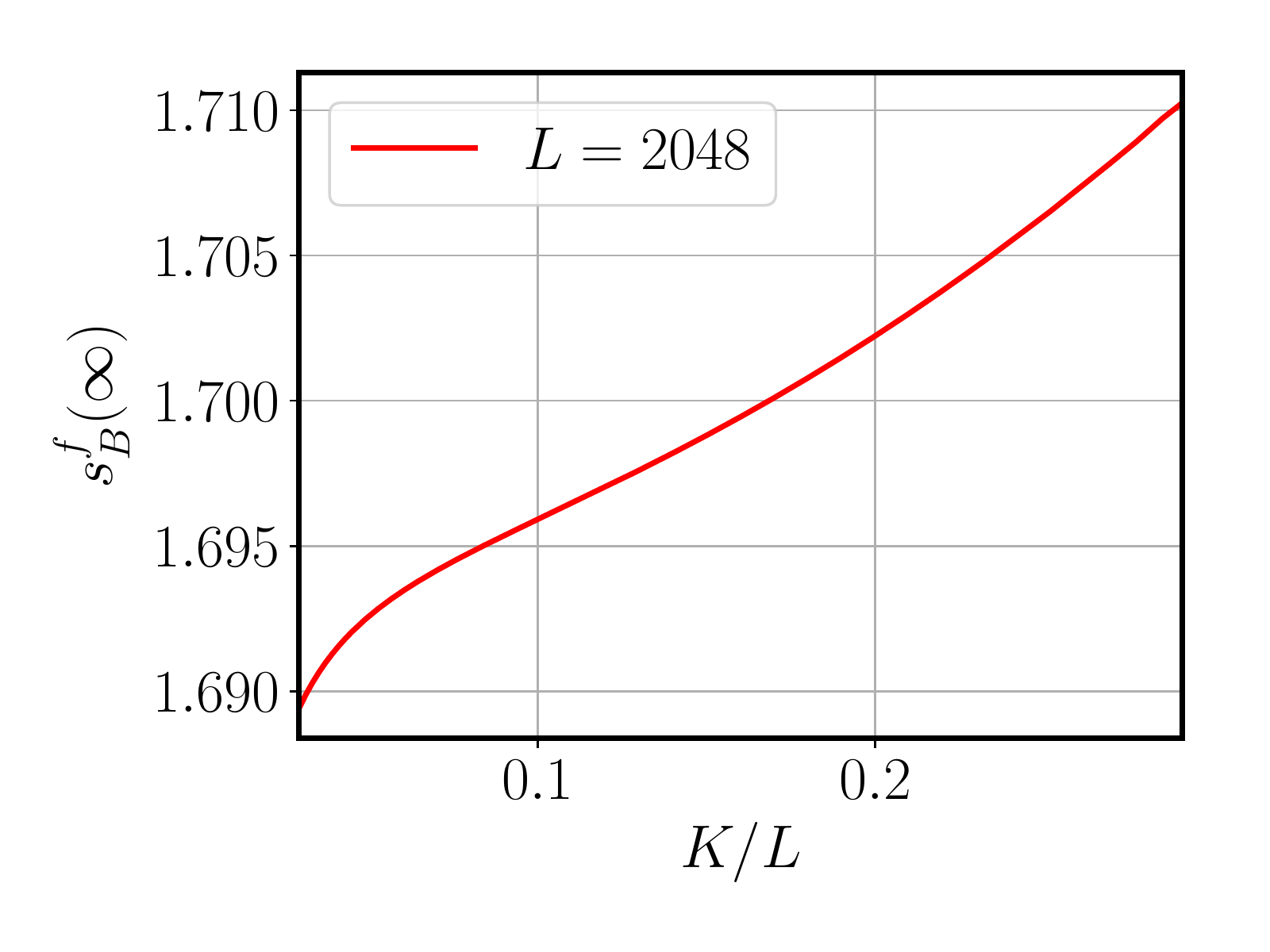}
}
\caption{{\bf Fermions}: The late time value of the f-macrostate entropy per particle, $s^f_B(\infty)$ as a function of the coarse-graining scale $K$ for the low and intermediate temperatures. It seems that we require $K /L \ll 1$ for the final value to not vary strongly with $K$. The variation also goes down with increasing temperature.} \label{sinf_fermions}
\centering

\subfigure[$T = 2 \pi/5$]{
	\includegraphics[scale=0.4]{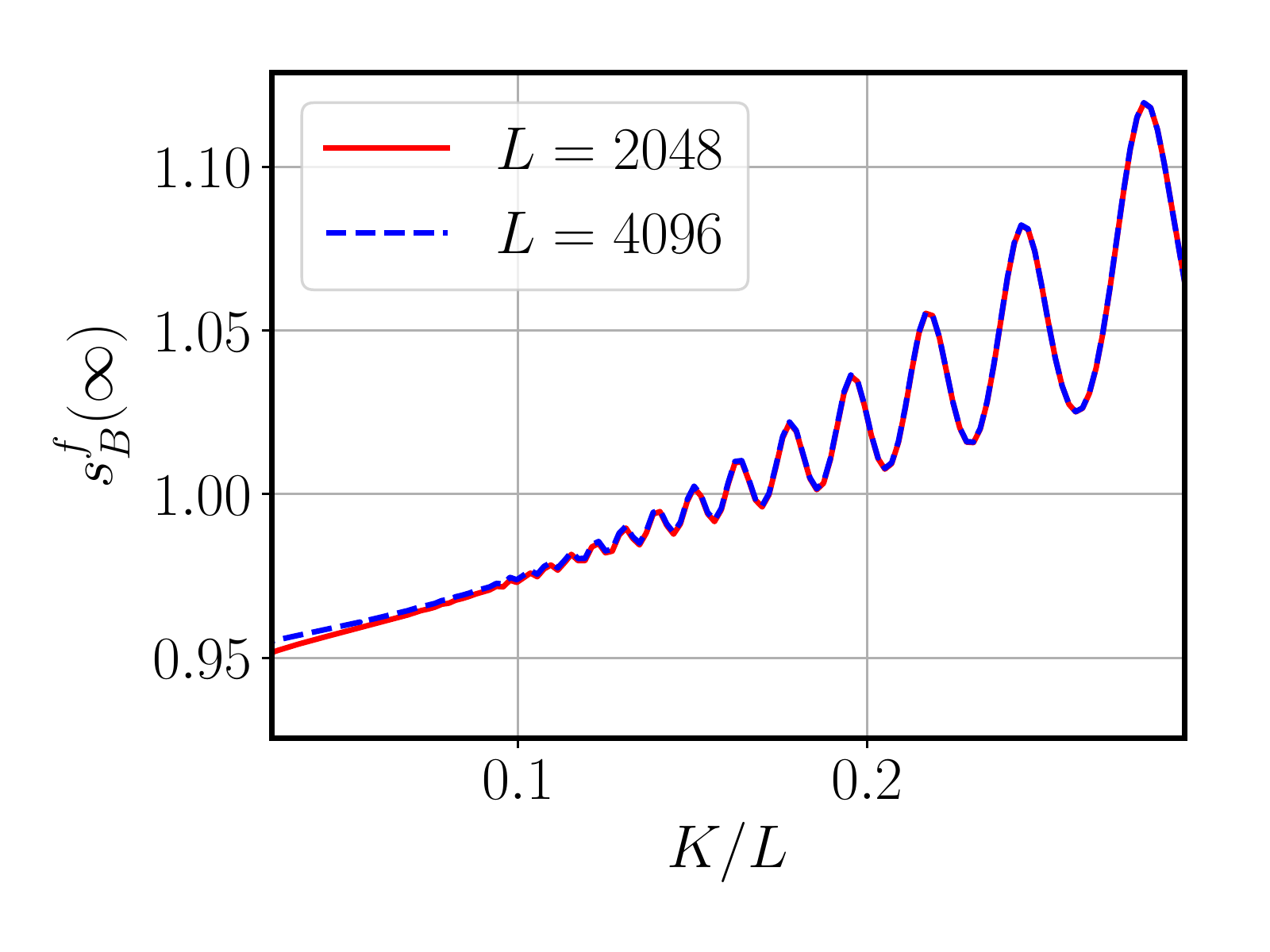}
}
\subfigure[$T = 2 \pi$]{
	\includegraphics[scale=0.4]{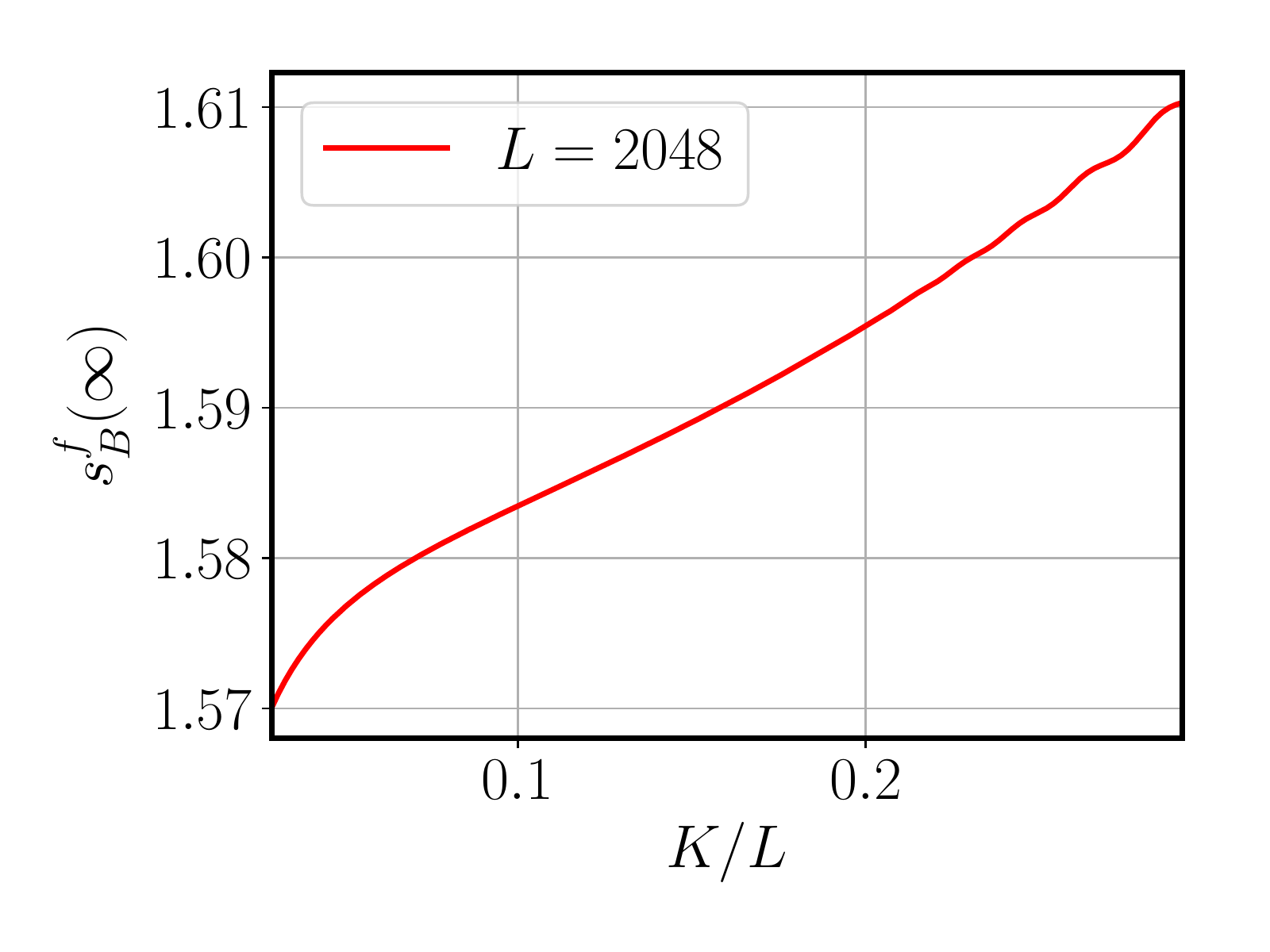}
}
\caption{{\bf Bosons}: The late time value of the f-macrostate entropy per particle, $s^f_B(\infty)$ as a function of the coarse-graining scale $K$ for the low and intermediate temperatures. It seems that we require $K /L \ll 1$ for the final value to not vary strongly with $K$. Also note that the variation with $K$ is stronger for bosons as compared to fermions, so a relatively smaller $K/L$ may be required for bosons. Here also, the variation goes down with increasing temperature.} \label{sinf_bosons}
\end{figure}

Here we briefly discuss the dependence of our results on the coarse-graining parameter $K$.

 {\bf Fermions}: In Fig.~(\ref{sinf_fermions}), we plot the saturation value $s_B^f(\infty)$ for different $K$ and for the low and intermediate temperatures. We observe that as long as $K \ll L$, the saturation value does not vary strongly with $K$. Also, the variation becomes smaller with increasing temperature.

{\bf Bosons}: The strong dependence on K can be understood from  Fig.~\eqref{sinf_bosons} which shows the variation of the saturation value of the f-macrostate entropy per particle with $K$. Here, we can clearly see that the variation is stronger for bosons as compared to fermions. As a result, the $K$ values used in our numerical study are not small enough compared to the system size $L$ for bosons, especially in the low-temperature case.

\section{Glossary}{\label{app:glossary}}

\begin{itemize}
    \item $f, U$: denote the two macrovariables studied in this work. 
    \item $S_B^f, S_B^U$: the Boltzmann entropies corresponding to the two macrovariables.
    \item $s_B^f, s_B^U$: Boltzmann entropies per particle.
    \item $X$: Phase space point.
    \item $\ket{\Phi}$: Pure state wavefunction.
    \item $N, E, L$: The total number of particles, the total energy, and the system size (Circle perimeter) respectively.
    \item $T, \beta, \mu$: The temperature, the inverse temperature, and the chemical potential respectively.
    \item $a$: Fraction of the circle occupied by the gas at time $t=0$. In most cases $a=1/2$.
    \item $m$: Mass of the particles.
    \item $\mathcal{H}$: Hilbert space.
    \item $\mathcal{H}_E$: Subspace of the full Hilbert space corresponding to the energy $E$.
    \item $\hat{M}_k$: Generic self-adjoint operators.
    \item $\nu, \mathcal{H}_{\nu}$: Generic macrostate and the corresponding macro-space.
    \item $|\mathcal{H}_{\nu}|$: Denotes the dimension of the $\mathcal{H}_{\nu}$ macro-space.
    \item $\hat{P}_{\nu}$: Projection operator onto the $\mathcal{H}_{\nu}$ macro-space.
    \item $S_{\nu}$: Boltzmann entropy corresponding to the $\nu$ macrostate.
    \item $\mathcal{H}_{eq}$: Equilibrium macro-space.
    \item $\hat{\rho}_{\text{GC}}, Z_{\text{GC}}$: Generalized canonical density operator and the corresponding partition function.
    \item $\lambda_k$: Lagrange multipliers (for satisfying the constraints) in the expression for $\hat{\rho}_{\text{GC}}$.
    \item $S_{\text{GvN}}$: The Gibbs-von Neumann entropy of the system
    \item $\delta, A$: Denote the size and number respectively of the coarse-grained cells in the $U$-macrostate. 
    \item $\ell$: Labels the coarse-grained cells in the $U$-macrostate. 
    \item $\hat{N}_{\ell}, \hat{P}_{\ell}, \hat{E}_{\ell}$: Denote the particle number, the momentum, and the energy operators respectively corresponding to the $\ell^{\text{th}}$ cell in the $U$-macrostate. 
    \item $K$: Denotes the coarse-graining scale in the $f$-macrostate.
    \item $\ket{\psi_{\alpha}} \equiv \ket{r, v}$: The wavepacket basis state localized in position around $x = r L/K$ and in momentum around $p = 2 \pi \hbar v/L$.
    \item $\hat{n}_{\alpha}, D_{\alpha}(t) \equiv D(r, v, t)$: Denote the occupation number operator and the average occupancy (wavepacket density) respectively of the wavepacket basis state $\ket{\psi_{\alpha}}$.
    \item $\ket{\chi_s}, e_s, n_s$: Denote the box energy eigenfunction, eigenvalue, and occupancy of the $s^{\text{th}}$ level respectively.
    \item $\ket{\varphi_n}, p_n, \epsilon_n$: Denote the circle energy eigenfunction, momentum, and energy eigenvalue of the $n^{\text{th}}$ level respectively.
    \item $\hat{\Psi}_x, \hat{b}_n$: Denote the annihilation operator in the position and in momentum space respectively.
    \item $\hat{\rho}_N$: The $N$-particle density operator.
    \item $\hat{\rho}_1, \rho_1(x, x'), \tilde{\rho}_1 (p_m, p_n)$: The single-particle density operator and its matrix elements in the position and momentum space respectively.
    \item $V$: Transformation matrix from the box to the circle basis.
    \item $f(e_s, \beta, \mu)$: Denotes the Fermi/Bose function.
    \item $\hat{\rho}_1^P, \hat{\rho}_1^M$: Denote the single-particle density operator corresponding to the pure and mixed state initial conditions respectively.
    \item $P(\{ n_s \})$: Denotes the grand canonical probability distribution of the configuration $\{ n_s \}$.
    \item $w(x, p, t)$: The Wigner function on the real line.
    \item $q_n$: Denotes the half-integer momenta on the circle.
    \item $w(x, q_n, t)$: The Wigner function on the circle.
    \item $\hat{N}, \hat{P}, \hat{E}$: Denote the total particle number, the total momentum, and the total  energy operators respectively.
    \item $\hat{n}(x, t), \hat{p}(x, t), \hat{e}(x, t)$: Denote the local particle, momentum, and energy density operators respectively.
    \item $n(x, t), p(x, t), e(x, t)$: Denote the expectation values of the local particle, momentum, and energy density operators respectively.
    \item $\mathcal{R}_v$: The set of $K$ integers centered around $v$.
    \item $v_{\pm}$: The two end points of the set $\mathcal{R}_v$. 
    \item $G_K(q, x)$: The localized kernel that, when integrated over the Wigner function, yields the wavepacket density.
    \item $D_r(r, t), D_v(v,t)$: The respective marginals of $D(r, v, t)$.
    \item $h_K(x)$: The localized kernel that, when integrated with the particle density $n(x, t)$, yields the coarse-grained marginal $D_r(r, t)$.
    \item $\hat{\rho}_N^{\star}$: Maximal N-particle density operator subject to the wavepacket density constraints. 
    \item $\lambda_{\alpha}$: Lagrange multipliers (for satisfying the constraints) in the expression for $\hat{\rho}_N^{\star}$.
    \item $\Delta s_B^f, \Delta s_B^U$: The final change in the entropies of $f$ and $U$ macrostates respectively. 
    \item $\rho$: Particle density $N/L$.
    \item $\lambda_{\text{th}}$: The thermal De-Broglie wavelength. 
    \item $\tau_p$: The period of oscillations of the particle density.
    \item $v_{\rm f}$: Fermi velocity. 
    \item $\tau$: The period of oscillations of the Boltzmann entropy $S_B^f$.
    \item $\tau_{rec}$: The recurrence period.
    \end{itemize}

\bibliographystyle{ieeetr}
\bibliography{bibliography.bib}

\begin{thebibliography}{10}

\bibitem{boltzmann1897}
L.~Boltzmann, ``On {Z}ermelo's paper ``{O}n the mechanical explanation of
  irreversible processes",'' {\em Annalen der Physik}, vol.~60, pp.~392--398,
  1897.

\bibitem{feynman2017character}
R.~Feynman, {\em The Character of Physical Law, with new foreword}.
\newblock MIT press, 2017.

\bibitem{lanford1976derivation}
O.~E. Lanford, ``On a derivation of the boltzmann equation,'' {\em
  Ast{\'e}risque}, vol.~40, no.~117, pp.~0353--70020, 1976.

\bibitem{penrose1990emperor}
R.~Penrose and N.~D. Mermin, ``The emperor’s new mind: Concerning computers,
  minds, and the laws of physics,'' 1990.

\bibitem{greene2004fabric}
B.~Greene, {\em The fabric of the cosmos: Space, time, and the texture of
  reality}.
\newblock Knopf, 2004.

\bibitem{lebowitz1993macroscopic}
J.~L. Lebowitz, ``Macroscopic laws, microscopic dynamics, time's arrow and
  boltzmann's entropy,'' {\em Physica A: Statistical Mechanics and its
  Applications}, vol.~194, no.~1-4, pp.~1--27, 1993.

\bibitem{Lebowitz_PT1993}
J.~L. Lebowitz, ``{B}oltzmann’s entropy and time’s arrow,'' {\em Physics
  Today}, vol.~46, no.~9, p.~32, 1993.

\bibitem{Goldstein_PD2004}
S.~Goldstein and J.~L. Lebowitz, ``On the ({B}oltzmann) entropy of
  non-equilibrium systems,'' {\em Physica D}, vol.~193, no.~1, p.~53, 2004.

\bibitem{griffiths1994}
R.~Griffiths, ``Statistical irreversibility: classical and quantum.,'' {\em
  Physical origins of time asymmetry}, pp.~147--159, 1994.

\bibitem{goldstein2017}
S.~Goldstein, D.~A. Huse, J.~L. Lebowitz, and R.~Tumulka, ``Macroscopic and
  microscopic thermal equilibrium,'' {\em Annalen der Physik}, vol.~529, no.~7,
  p.~1600301, 2017.

\bibitem{tasaki2016}
H.~Tasaki, ``Typicality of thermal equilibrium and thermalization in isolated
  macroscopic quantum systems,'' {\em Journal of Statistical Physics},
  vol.~163, no.~5, pp.~937--997, 2016.

\bibitem{mori2018}
T.~Mori, T.~N. Ikeda, E.~Kaminishi, and M.~Ueda, ``Thermalization and
  prethermalization in isolated quantum systems: a theoretical overview,'' {\em
  Journal of Physics B: Atomic, Molecular and Optical Physics}, vol.~51,
  no.~11, p.~112001, 2018.

\bibitem{de2006}
W.~De~Roeck, C.~Maes, and K.~Neto{\v{c}}n{\`y}, ``Quantum macrostates,
  equivalence of ensembles, and an h-theorem,'' {\em Journal of mathematical
  physics}, vol.~47, no.~7, p.~073303, 2006.

\bibitem{callen1998}
H.~B. Callen, ``Thermodynamics and an introduction to thermostatistics,'' 1998.

\bibitem{chakraborti2021entropy}
S.~Chakraborti, A.~Dhar, S.~Goldstein, A.~Kundu, and J.~L. Lebowitz, ``Entropy
  growth during free expansion of an ideal gas,'' 2021.

\bibitem{dean2018fermions}
D.~S. Dean, P.~Le~Doussal, S.~N. Majumdar, and G.~Schehr, ``Wigner function of
  noninteracting trapped fermions,'' {\em Phys. Rev. A}, vol.~97, p.~063614,
  Jun 2018.

\bibitem{manas2018}
M.~Kulkarni, G.~Mandal, and T.~Morita, ``Quantum quench and thermalization of
  one-dimensional fermi gas via phase-space hydrodynamics,'' {\em Phys. Rev.
  A}, vol.~98, p.~043610, Oct 2018.

\bibitem{eisler2013}
V.~Eisler and Z.~R{\'a}cz, ``Full counting statistics in a propagating quantum
  front and random matrix spectra,'' {\em Physical review letters}, vol.~110,
  no.~6, p.~060602, 2013.

\bibitem{scopa2021}
S.~Scopa, A.~Krajenbrink, P.~Calabrese, and J.~Dubail, ``Exact entanglement
  growth of a one-dimensional hard-core quantum gas during a free expansion,''
  {\em Journal of Physics A: Mathematical and Theoretical}, vol.~54, no.~40,
  p.~404002, 2021.

\bibitem{rigol2011}
L.~F. Santos, A.~Polkovnikov, and M.~Rigol, ``Entropy of isolated quantum
  systems after a quench,'' {\em Phys. Rev. Lett.}, vol.~107, p.~040601, Jul
  2011.

\bibitem{deutsch2019a}
D.~{\v{S}}afr{\'a}nek, J.~M. Deutsch, and A.~Aguirre, ``Quantum coarse-grained
  entropy and thermodynamics,'' {\em Physical Review A}, vol.~99, no.~1,
  p.~010101, 2019.

\bibitem{deutsch2019b}
D.~{\v{S}}afr{\'a}nek, J.~Deutsch, and A.~Aguirre, ``Quantum coarse-grained
  entropy and thermalization in closed systems,'' {\em Physical Review A},
  vol.~99, no.~1, p.~012103, 2019.

\bibitem{solano2016theory}
E.~Solano-Carrillo and A.~Millis, ``Theory of entropy production in quantum
  many-body systems,'' {\em Physical Review B}, vol.~93, no.~22, p.~224305,
  2016.

\bibitem{fick1990}
E.~Fick, G.~Sauermann, and W.~D. Brewer, {\em The quantum statistics of dynamic
  processes}, vol.~86.
\newblock Springer, 1990.

\bibitem{wigner1932quantum}
E.~Wigner, ``On the quantum correction for thermodynamic equilibrium,'' {\em
  Phys. Rev.}, vol.~40, pp.~749--759, Jun 1932.

\bibitem{hillery1984}
M.~Hillery, R.~F. O'Connell, M.~O. Scully, and E.~P. Wigner, ``Distribution
  functions in physics: Fundamentals,'' {\em Physics reports}, vol.~106, no.~3,
  pp.~121--167, 1984.

\bibitem{mukunda1978algebraic}
N.~Mukunda, ``Algebraic aspects of the wigner distribution in quantum
  mechanics,'' {\em Pramana}, vol.~11, no.~1, pp.~1--15, 1978.

\bibitem{mukunda1979wigner}
N.~Mukunda, ``Wigner distribution for angle coordinates in quantum mechanics,''
  {\em American Journal of Physics}, vol.~47, no.~2, pp.~182--187, 1979.

\bibitem{berry1977}
M.~V. Berry, ``Semi-classical mechanics in phase space: a study of wigner’s
  function,'' {\em Philosophical Transactions of the Royal Society of London.
  Series A, Mathematical and Physical Sciences}, vol.~287, no.~1343,
  pp.~237--271, 1977.

\bibitem{dhar2022}
A.~Dhar {\em ,~to be published}.

\bibitem{husimi1940}
K.~Husimi, ``Some formal properties of the density matrix,'' {\em Proceedings
  of the Physico-Mathematical Society of Japan. 3rd Series}, vol.~22, no.~4,
  pp.~264--314, 1940.

\bibitem{wehrl1979}
A.~Wehrl, ``On the relation between classical and quantum-mechanical entropy,''
  {\em Reports on Mathematical Physics}, vol.~16, no.~3, pp.~353--358, 1979.

\bibitem{tsukiji2016}
H.~Tsukiji, H.~Iida, T.~Kunihiro, A.~Ohnishi, and T.~T. Takahashi, ``Entropy
  production from chaoticity in yang-mills field theory with use of the husimi
  function,'' {\em Physical Review D}, vol.~94, no.~9, p.~091502, 2016.

\bibitem{kunihiro2009}
T.~Kunihiro, B.~M{\"u}ller, A.~Ohnishi, and A.~Sch{\"a}fer, ``Towards a theory
  of entropy production in the little and big bang,'' {\em Progress of
  theoretical physics}, vol.~121, no.~3, pp.~555--575, 2009.

\bibitem{goes2020}
B.~O. Goes, G.~T. Landi, E.~Solano, M.~Sanz, and L.~C{\'e}leri, ``Wehrl entropy
  production rate across a dynamical quantum phase transition,'' {\em Physical
  Review Research}, vol.~2, no.~3, p.~033419, 2020.

\bibitem{dean2019}
D.~S. Dean, P.~Le~Doussal, S.~N. Majumdar, and G.~Schehr, ``Nonequilibrium
  dynamics of noninteracting fermions in a trap,'' {\em EPL (Europhysics
  Letters)}, vol.~126, no.~2, p.~20006, 2019.

\bibitem{chakraborti2023}
S.~Chakraborti, A.~Dhar, and A.~Kundu, ``Boltzmann’s entropy during free
  expansion of an interacting gas,'' {\em Journal of Statistical Physics},
  vol.~190, no.~4, p.~74, 2023.

\end{thebibliography}

\end{document}